\documentclass[sigconf]{acmart}
\AtBeginDocument{%
  }

\copyrightyear{2026}
\acmYear{2026}
\setcopyright{cc}
\setcctype{by}
\acmConference[CHI '26]{Proceedings of the 2026 CHI Conference on Human Factors in Computing Systems}{April 13--17, 2026}{Barcelona, Spain}
\acmBooktitle{Proceedings of the 2026 CHI Conference on Human Factors in Computing Systems (CHI '26), April 13--17, 2026, Barcelona, Spain}
\acmPrice{}
\acmDOI{10.1145/3772318.3790534}
\acmISBN{979-8-4007-2278-3/2026/04}

\usepackage{graphicx}
\usepackage{tikz}
\usepackage{hf-tikz}
\usepackage{enumitem}
\usepackage{todonotes}

\usepackage{xspace}

\newcommand{\systemname}{PrevizWhiz\xspace}

\newcommand\red[1]{\textcolor{black}{#1}}

\definecolor{Darkgreen}{rgb}{0.0, 0.5, 0.0}

\newcommand{\etal}{\emph{et al.}}

\newcommand{\eg}{\emph{e.g.}}

\usepackage{booktabs,tabularx}
\usepackage{multirow}
\usepackage{soul}
\usepackage{xspace}
\usepackage{cleveref}
\usepackage[utf8]{inputenc}
\usepackage{enumitem}
\usepackage{subcaption}
\usepackage{ifthen}
\usepackage{graphicx}
\usepackage{colortbl} 
\crefformat{section}{\S#2#1#3} 
\crefformat{subsection}{\S#2#1#3}
\crefformat{subsubsection}{\S#2#1#3}

\begin{document}

\title{
\systemname: 
Combining Rough 3D Scenes and 2D Video to Guide Generative Video Previsualization
}
\author{Erzhen Hu}
\email{eh2qs@virginia.edu}
\affiliation{%
  \institution{Autodesk Research \& University of Virginia}
  \city{Toronto}
  \state{Ontario}
  \country{Canada}
}

\author{Frederik Brudy}
\email{frederik.brudy@autodesk.com}
\orcid{0000-0002-3868-0967}

\affiliation{
 \institution{Autodesk Research}
  \city{Toronto}
  \state{Ontario}
  \country{Canada}
}

\author{David Ledo}
\email{david.ledo@autodesk.com}
\orcid{0000-0002-0455-7936}

\affiliation{
 \institution{Autodesk Research}
  \city{Toronto}
  \state{Ontario}
  \country{Canada}
}

\author{George Fitzmaurice}
\email{george.fitzmaurice@autodesk.com}
\orcid{0000-0002-2834-7757}

\affiliation{
 \institution{Autodesk Research}
  \city{Toronto}
  \state{Ontario}
  \country{Canada}
}

\author{Fraser Anderson}
\email{fraser.anderson@autodesk.com}
\orcid{0000-0003-3486-8943}

\affiliation{
\institution{Autodesk Research}
  \city{Toronto}
  \state{Ontario}
  \country{Canada}
}


\begin{abstract}

In pre-production, filmmakers and 3D animation experts must rapidly prototype ideas to explore a film’s possibilities before full-scale production, yet conventional approaches involve trade-offs in efficiency and expressiveness.
Hand-drawn storyboards often lack spatial precision needed for complex cinematography, while 3D previsualization demands expertise and high-quality rigged assets.
To address this gap, we present \systemname, a system that leverages rough 3D scenes in combination with generative image and video models to create stylized video previews. 
The workflow integrates frame-level image restyling with adjustable resemblance, time-based editing through motion paths or external video inputs, and refinement into high-fidelity video clips.
A study with filmmakers demonstrates that our system lowers technical barriers for film-makers, accelerates creative iteration, and effectively bridges the communication gap, while also surfacing challenges of continuity, authorship, and ethical consideration in AI-assisted filmmaking.
\end{abstract}

\ccsdesc[500]{Human-centered computing~Interactive systems and tools}
\ccsdesc[500]{Human-centered computing}

\keywords{previsualization, generative AI, filmmaking, human-AI collaboration, creativity support}


\begin{teaserfigure}  
\centering
    \includegraphics[width=\linewidth]{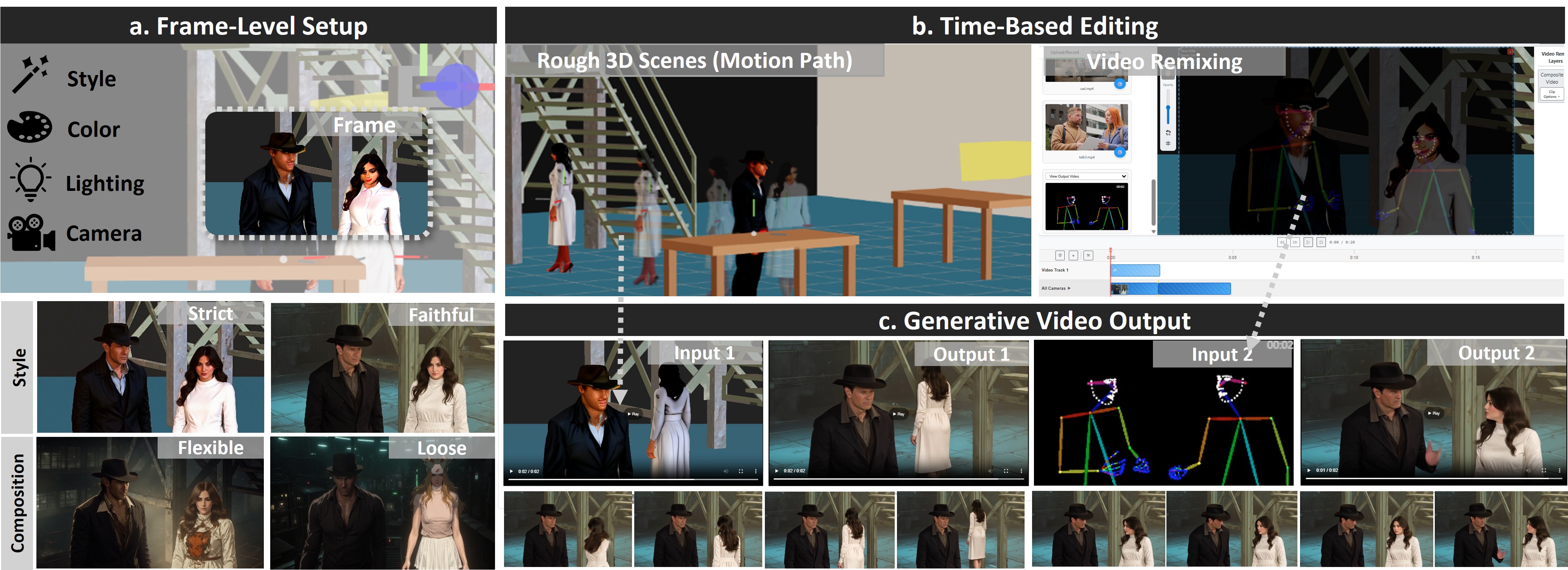}
    \caption{\systemname supports an authoring workflow that transforms rough 3D scenes and video remixing into generative videos. (a1) Users begin with a basic setup, capturing frame-level images from the 3D environment where color, lighting, and camera settings can be defined. (a2) Restyled images are added as inputs, with resemblance adjustable from strict to faithful, flexible, or loose, depending on how closely style, color, lighting, and spatial composition adhere to the original 3D setup.  (b) They can then perform time-based editing by animating elements on the timeline or importing external videos to define motion, which will be exported as the input to the video generation model; (c) Finally, we show the result of video output that use (c1) 3D scenes or (c2) the video library as a guidance for generative video models.
    }
    \label{fig:teaser}
\end{teaserfigure}

\maketitle

\section{Introduction}

Previsualization (previz) is a central practice in filmmaking, enabling directors and creative teams to explore the visual and narrative structure of a scene before production~\cite{wong2012previsualization}. By creating early visualizations, filmmakers can test ideas for camera angles, blocking, pacing, and emotional beats without the expense of full-scale sets, actors, or detailed assets. Beyond its role as a creative sketching tool, previz also functions as a collaborative artifact, helping directors, cinematographers, production designer, and other stakeholder align around a shared vision~\cite{10.1145/2207676.2207700,block2020visual,hart2013art}.

Despite its importance, existing approaches force filmmakers to make trade-offs between speed, fidelity, and control. Storyboards and moodboards are quick and expressive, allowing for early exploration and communication of creative intent~\cite{haesen2009supporting,10.1145/3313831.3376520}. However, these are static, offering limited spatial and temporal representation: they cannot adequately represent motion or timing, making it difficult to visualize complex shots or sequences. 3D previz tools on the other hand allow filmmakers to compose scenes, experiment with camera blocking, and ensure continuity across shots~\cite{wong2012previsualization,muender2018empowering, 8798181}. However, these tools require high-fidelity 3D assets, rigging, and animation expertise~\cite{rao2023dynamic}. Many existing 3D previz tools also fail to convey fine-grained nuances like emotional beats and micro-actions.

Recent advances in generative AI can accelerate previsualization by producing images or videos directly from textual prompts, allowing filmmakers to quickly generate outputs with a compelling visual style~\cite{10.1145/3706598.3713342}. Yet, they pose new challenges. Text-to-image and text-to-video models often struggle with temporal consistency, making coherent motion across frames challenging~\cite{pal2025illuminating}. They also lack spatial grounding: precise placement of objects and camera, blocking, and continuity are difficult to control. As a result, current approaches using generative AI risk producing highly polished looking results that are disconnected from the filmmaker's intended structure. Filmmakers need a lightweight and flexible approach that combines the spatial grounding of 3D tools with the expressive richness of generative video tools. 


We present \systemname, a system that allows filmmakers to rapidly explore and visualize their shots by combining rough 3D scene blocking for timing and spatial structure, 2D video references for detailed character motion, and generative stylization guided by images and text. 
Filmmakers begin by arranging rough 3D proxies to establish prop positions, character movement, as well as camera paths (\autoref{fig:teaser}a). They can restyle frames from their 3D scenes to experiment with different aesthetic styles, ranging from strict adherence to loose reinterpretation of their compositions (\autoref{fig:teaser}a). 
Finally, \systemname allows filmmakers to specify three levels of motion fidelity: (1) coarse motion from 3D blocking, (2) stylized animations that combine motion from 3D blocking with the restyled frame, and (3) control-video animation that augments the stylized animation with 2D reference videos for detailed character motion (\autoref{fig:teaser}b). 
These frames of scene composition, and time-based elements can guide the video generation of final outputs (\autoref{fig:teaser}c), shaping style, lighting, composition, and movement in ways that balance 
the structural consistency of 3D blocking with the expressiveness of 2D generative tools to create previsualization for film.

Our contributions are: (1) \systemname, a system that combines rough 3D blocking, frame stylization, and granular animation control to enable lightweight yet expressive previsualization, and (2) findings from a user study with filmmakers and 3D artists showing how the system enables rapid ideation during previsualisations and probes their thoughts on generative tools for pre-production. 

\section{Related Work}
\label{sec:relatedwork}
Our work builds on prior research in previz practices, AI and generative tools for previz, and generative approaches for style transfer and guided video generation.
\subsection{Pre-Production and Previz Tools During Filmmaking}
Previz techniques help filmmakers experiment with shots, blocking, pacing, and mood before committing to full-scale production~\cite{wong2012previsualization}. Conventional approaches such as storyboards 
and moodboards are fast and expressive but remain static~\cite{haesen2009supporting,10.1145/3313831.3376520}, offering limited spatial or temporal fidelity. 

In contrast, 3D previz tools~\cite{muender2018empowering} offer precise composition and blocking, but they demand significant expertise, asset preparation, time, and budget, and are usually used in filming teams with larger budget and scale~\cite{wong2012previsualization}. Test videos~\cite{10.1145/2556288.2557304,10.1145/3613904.3642575} reduce barriers by mimicking live filming, but rely heavily on physical spaces, making iteration tedious (\eg, reshooting actions for each camera angle). Filmmakers thus face a trade-off between speed, cost, expressiveness, and fidelity when it comes to previz. Our system draws inspiration from these practices, combining the speed of rough sketches and storyboards, the detail of 3D tools, and the expressiveness of control videos such as online videos and test video footage.

Previz also serves as a cross-disciplinary communication tool bridging directors, cinematographers, art, sound, and animation teams~\cite{10.1145/2207676.2207700,block2020visual,hart2013art}. While tools such as SyncSketch \footnote{\url{https://syncsketch.com/}} and Autodesk Flow Capture \footnote{\url{https://www.moxion.io/}} allow annotation of images and videos for collaboration, there remains a need for previz that looks more “finished” while still supporting rapid iteration. 

Our work builds on this line of  previz research and commercial tools by using 3D environments not as \textit{finished high-fidelity assets}, but as a shared rough environment for cross-disciplinary discussion, and rapid iteration. By combining these environments with generative 2D video outputs, we enable lightweight yet expressive previz that supports both rough and polished states, aligning composition, lighting, style, and performance without requiring deep 3D expertise.

\subsection{AI and Generative Tools For Previz}


Researchers have explored AI-based tools to support story ideation and visualization beyond hand-drawn sketches. Early systems such as Schematic Storyboarding~\cite{goldman2006schematic} and dynamic storyboards~\cite{rao2023dynamic} generated static and animated frames from scripts and imported sets. More recent approaches leverage data retrieval and generative models. For example, ScriptViz~\cite{10.1145/3654777.3676402}
retrieves images from movie datasets based on script attributes (\eg, location, time of day, characters), allowing filmmakers to visualize their script on-the fly. Generative methods have been used to restyle images, such as  CineVision~\cite{wei2025cinevision} which combines scriptwriting with retrieved movie references to support director–cinematographer collaboration. Commercial tools like Previs Pro\footnote{\url{https://www.previspro.com/}} enable static storyboard capture from 3D scenes with limited style adjustments. In contrast, our work emphasizes user-authored 3D scenes rather than database-driven retrieval. These scenes can then be restyled into polished outputs, preserving creative intent while enabling flexible and rapid iteration. 


Commercial and research tools extend beyond storyboards by incorporating video. For instance, DigitalFish\footnote{\url{https://www.digitalfish.com/onsight/}} and Jetset\footnote{\url{https://lightcraft.pro/jetset/}} provide real-time compositing of live footage with 3D environments, while volumetric capture~\cite{10488869} allows actors to perform inside virtual backdrops. CollageVis~\cite{10.1145/3613904.3642575} supports 2.5D video composition through segmentation and recomposition, while RADiCAL\footnote{\url{https://radicalmotion.com/}} offers high-fidelity motion capture and 3D previz, but demands heavy asset pipelines. These systems are either static (storyboards/moodboards), database-driven (retrieval), or high-cost (volumetric/3D assets). By contrast, \systemname blends rough 3D setups with 2D video references and integrates generative video models, producing expressive yet structured outputs for previz. Notably, prior work has not systematically integrated generative video models~\cite{10.1145/3698061.3726926} into expressive previz workflows.

\subsection{Generative Approaches: Style Transfer and 
Guided Generation}

\subsubsection{AI Adoption in Filmmaking Workflows}
Recent work has demonstrated a shift in filmmaking driven by new workflows with AI: Freeman \etal~\cite{10.1145/2556288.2557304} proposed live, tablet-based editing during production; Halperin \etal~\cite{10.1145/3706598.3713342} examined how amateurs use generative AI in filmmaking courses; and Chung \etal~\cite{10.1145/3706598.3713532} introduced AI editing assistants for live broadcasting. Anderson and Niu~\cite{10.1145/3706599.3719991} analyzed YouTube tutorials, finding most generative AI use occurs in post-production (e.g., restyling, VFX, upscaling). 
Our system differs by bringing generative video earlier into pre-production, enabling directors and cinematographers to iterate before post-production. 

\subsubsection{Image Generation: Style Transfer and Flow-Based Methods}
Recent approaches have explored multiple strategies for script-to-visual generation.
\red{Some recent approaches such as GANs~\cite{karras2019style,zhu2017unpaired,isola2017image} and VAEs attempt to generate visual scenes directly from scripts, combining script analysis with image retrieval to align textual descriptions with visual assets.} 
While effective in constrained domains, such systems~\cite{10.1145/3654777.3676402} often rely heavily on pre-defined templates, large-scale movie datasets, or visual writing prompts, which limits flexibility for creating novel or stylistically unique visuals.

In contrast, text-to-image models (\eg, Stable Diffusion~\cite{rombach2022high}, DALL·E~\cite{ramesh2021zero}, Imagen~\cite{saharia2022photorealistic}) generate entirely new images directly from textual prompts, offering much greater creative freedom. However, they often lack fine-grained control and structural consistency across scenes. Style transfer methods address the opposite problem: they preserve the spatial composition of an input image while altering its appearance according to a target style (e.g., anime, film noir)~\cite{wei2025illuminating}. Yet, their scope is narrower, as they cannot easily recompose or invent new elements beyond the given input.
Flow-based editing techniques such as FlowEdit~\cite{kulikov2024flowedit} extend beyond classical style transfer by transporting features from a source distribution to a target distribution while maintaining structural correspondences. This allows more flexible transformations that balance content preservation with stylistic change. Combined with spatial conditioning techniques like ControlNet~\cite{zhang2023adding}, such methods support controllable resemblance to the 3D scenes while enabling selective restyling.

Building on these advances, \systemname integrates generative editing with the structural benefits of maintaining spatial composition from 3D setups. With Flow Edit and ControlNet methods, \systemname enables the controllability of colour, lighting, and style to the degree that filmmakers need to realize their vision.
Filmmakers can also determine which aspects of a scene—such as spatial composition, movement, colour, or lighting—are preserved from the 3D environment and which are reimagined through generative models.

\subsubsection{Driving Movements With Multi-Modal Methods}
Motion in previsualization has been added to scenes via 3D motion libraries (\eg, Mixamo\footnote{\url{https://www.mixamo.com/}}
), text-to-motion retrieval~\cite{petrovich2023tmr},  text-to-motion generative models~\cite{guo2022generating}, and pose estimation methods for video-to-motion~\cite{10.1145/3567728,sun2021monocular,fang2017rmpe} pipelines. While effective for prototyping, they struggle with subtle gestures, expressions, and multi-character dynamics essential for storytelling~\cite{gao2024identity,dhall2017individual}. 

Recent advances in generative video models have introduced new opportunities for video creation, though challenges remain around control, continuity, and integration into production workflows. Text-to-image models such as Stable Diffusion~\cite{rombach2022high} enable flexible styling with spatial guidance (e.g., ControlNet~\cite{zhang2023adding}), and extensions such as Stable Video Diffusion~\cite{blattmann2023stable} or commercial systems like Sora and Veo 3 extend these capabilities to video. However, issues of temporal coherence and fine-grained controllability limit their applicability in production. Furthermore, despite their impressive results, these interactive systems and control modalities still face substantial
limitations, and the intuitiveness of the control modalities and formats, particularly when it comes to complex design tasks.

Recent systems such as Wan Fun Control~\cite{wan2025} and VACE~\cite{jiang2025vace} add multi-modal guidance (e.g., skeleton, depth, line art), yet remain difficult to apply in structured filmmaking due to the speed, and limited duration of video generated. Applying these recent generative video methods, 
our approach integrates 3D-defined whole-body motions with 2D video-based gesture/expression controls, creating a hybrid video guidance method that balances macro-level blocking with micro-level expressiveness. 

\section{\systemname}
\systemname is a system that allows filmmakers to create previsualizations by combining 3D scene blocking, stylization using generative AI, and detailed character motion driven by video references. 

\subsection{Design Rationale}
Our design rationale is centered around lowering the barriers for filmmakers to rapidly create previsualizations without requiring high-fidelty 3D assets or advanced technical expertise. During conception of \systemname, and informed by prior literature (Section~\ref{sec:relatedwork}), we identified three key aspects of the system: 1) using rough 3D scenes to block out the spatial environment, 2) enabling frame stylization for refined previews, and 3) providing rapid means for creating motion through varying levels of fidelity. 

\paragraph{\textit{R1. Support scene setup with rough 3D blocking}} 
Filmmakers use previsualization to explore the spatial and temporal structure of scenes. Conventional previz workflows often rely on high-fidelty 3D assets and skilled animators, which slows down early exploration. However, prior work has shown the value of rough or low-fidelity input to support rapid iteration in creating previsualizations \cite{10.1145/3613904.3642575,rao2023dynamic}.
Inspired by these prior works and low-fi sketching practices, we use rough 3D environments to anchor generative outputs while keeping iteration lightweight. These rough scenes encode spatial and temporal cues (e.g., character positions, camera blocking) without requiring polished assets, accelerating experimentation. 

\paragraph{\textit{R2. Preview style for iteration and communication}}
Previsualization is not only about blocking and motion, but also about communicating tone and aesthetics to collaborators and stakeholders. Prior work around storyboards \cite{chen2019neural}, moodboards \cite{10.1145/3532106.3533565,10.1145/3563657.3596131}, and previsualization \cite{wei2025cinevision} highlights the importance of quickly exploring visual style. 
We therefore allow filmmakers to stylize captured frames from the 3D environment into different looks. By letting users set how strict or loose the restyled image should adhere to composition and style, we bridge between rough environments and envisioned cinematic aesthetic. 
As a result, filmmakers can not only iterate on their desired look, but can also communicate more clearly with others. 

\paragraph{\textit{R3. Support multiple levels of fidelity}}
Previsualization varies in how much precision, refinement and control is needed. Sometimes coarse, blocky motion is enough to test pacing, while at other times granular motion such as detailed gestures and emotional facial expressions are needed. To address this, our workflow provides three levels of control:
\begin{enumerate}[leftmargin=*]
    \item \textit{Motion from 3D blocking}, conveys timing and positioning directly within the 3D scene. Rough character (i.e., simply shifting an avatar's position without any motion) and camera movements are established in the 3D environment by adjusting their placement and orientation in the space. 
    \item \textit{Stylized motion} generates a video by combining motion from 3D blocking with the restyled frame that defines the target visual style. Coarse character translations are turned into more detailed movement, incorporating body kinematics while applying a consistent desired style. 
    \item \textit{Control-video motion} augments stylized motion with fine-grained motion by \red{leveraging external video references such as online videos or test video footages}. These references are layered on top of the 3D blocking to serve as templates for detailed movements, such as gestures or facial expressions. 
\end{enumerate}

\subsection{Interface Overview}
\systemname{} is web application that is comprised of several panels of independent functionality (\autoref{fig:basic panels}, \autoref{fig:layers}). The Scene Blocking panel (\autoref{fig:basic panels}b-c) allows the filmmaker to configure the 3D environment of their scene, and reposition the characters, cameras and elements within the scene. It also contains an editable timeline that allows for adding and modification of keyframes to support the rough animation of characters and cameras. The Image Styling and Animation panel allows filmmakers to use text descriptions and image references to apply a visual treatment to the shots that are recorded from the playback of the 3D scene (\autoref{fig:basic panels}d). Lastly, the Granular Motion Control Panel allows filmmakers to add and modify videos that provide motion references to further modify the motion of characters in the scene, for example by adding facial expressions and gestures to the characters (\autoref{fig:layers}).

\subsection{Scene Blocking and Composition}
\label{sec:basic}

\begin{figure*}[ht]
    \centering
    \includegraphics[width=\linewidth]{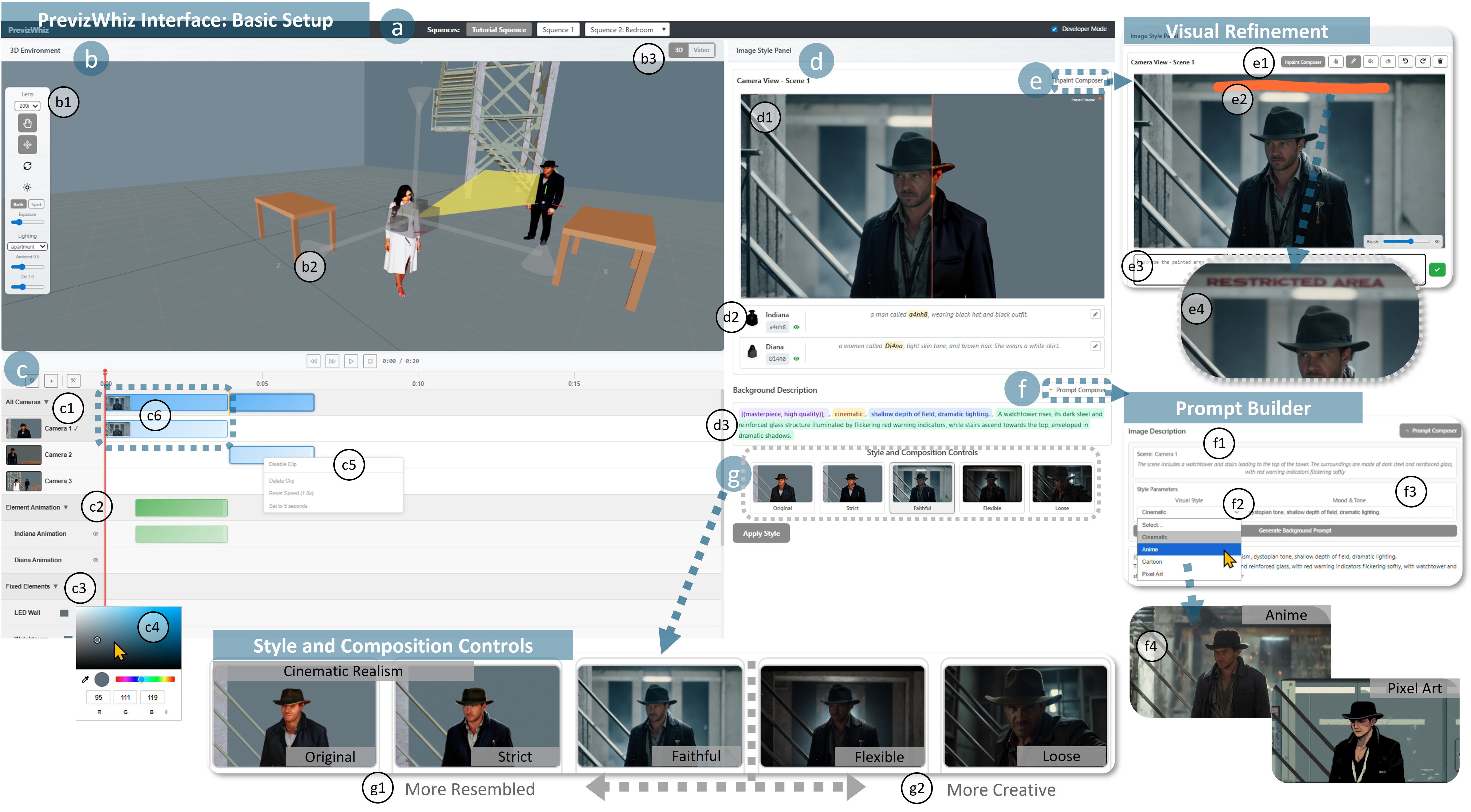}
    \caption{\systemname Scene Blocking and Composition Overview: (a) where users can select different squences and scenes. (b)
    3D Environment Panel for setting up cameras (b1), lens and lighting, and (b2) exploring the 3D scene pan/tilt/orbit controls.  
    (c) Timeline Panel for blocking camera, avatar, and element movements, including (c1) camera tracks, (c2) element animations, (c3) fixed/movable elements, (c4) color adjustments, and (c5) clip editing; (c6) the clip with restyled images attached.
    (d) Image Style Panel where (d1) 3D input and 2D output can be compared, with the prompt inputs including the (d2) description of customized characters, and (d3) background prompts entered or composed.
    (e) Inpainting composer tools with (e1) brush tools on identifying (e2) editable regions, (e3) text prompts describing the target details, which can help getting (e4) applied details such as painting a stenciled text.
    (f) Prompt Composer with (f1) basic background scene descriptions, (f2) visual style options, and (f3) mood/tone settings. 
    (g) Resemblance control to balance resemblance vs. creativity in the generated output.
    \red{Full video and scene output of this scene is shown in Appendix.}
        }
    \label{fig:basic panels}
\end{figure*}

The \textbf{Scene Blocking and Composition panel} allows the filmmaker to define the (1) \textit{scene composition} (\autoref{fig:basic panels}b) such as spatial layout with basic shapes, colours and lighting of the scene as well as (2) \textit{time-based elements} (\autoref{fig:basic panels}c) such as the rough camera tracking/blocking and creating motion paths of elements. The cameras placed in this scene are used to record the clips that are later processed by the generative models to render the final result.

\subsubsection{3D Environment and Timeline}

\paragraph{3D Scene Manipulation}
Accurate composition of the shot is important to filmmakers, and \systemname{} provides a 3D environment that enables them to position objects within the space. Characters, objects, lights and cameras can be placed within the scene and moved using standard 3D manipulation gizmos (\autoref{fig:basic panels}b1-2). The user can use standard pan/tilt/orbit controls to navigate the primary viewport of the scene, along with the camera preview to view the scene from the active camera.

By changing lighting (\autoref{fig:basic panels}b1) and colouring (\autoref{fig:basic panels}c4) of objects, the user can add more control to the look and feel of the shot during scene composition. Changing these parameters can better reflect the on-set appearance, or the practicalities of lighting a particular set. The impact of colouring and lighting can be controlled in the later stylization stage by specifying the system's adherence to the reference images (\autoref{fig:resemble-vs-decouple} and \autoref{fig:levels of adherence}).


\subsubsection{Animating Cameras and Objects}

The timeline is comprised of three main tracks that separate the key objects in the scene: (1) camera tracks (\autoref{fig:basic panels}c1), which allow users to create and edit cameras that will be used to frame and capture the scene; (2) animation tracks (\autoref{fig:basic panels}c2), which display motion paths of movable objects such as characters or vehicles; and (3) fixed element tracks (\autoref{fig:basic panels}c3), which represent static components of the scene which do not move, but have parameters that might change over time (\eg, colour, lighting properties). These tracks establish a clear hierarchy of scene elements, enabling users to easily navigate between the different objects and cameras. 

Pre-defined cameras are displayed as a list of thumbnails (\autoref{fig:basic panels}c1), allowing users to switch views directly. These pre-defined cameras represent common lenses and parameters used in filmmaking and allow the user to quickly frame and understand their scene. Additional cameras or motion-enabled elements can be added with the + button, with their actions recorded in the corresponding track.

\paragraph{Moving Cameras and Objects} 
Movable elements can be animated by recording motion paths directly in the 3D scene, either through keyboard input (WASD + Q/E) or by dragging elements with the 3D translation gizmo. These rough trajectories, such as walking, running, or jumping, are then represented on the Element Animation Track (\autoref{fig:basic panels}c2) as clips. These rough movements provide structured motion paths that serve as one of the key inputs for the generative video model, ensuring that the final output reflects both spatial and temporal continuity.

Camera motion can also be defined by specifying start and end keyframes for position, rotation, and lens, with interpolation handled automatically, similar to conventional 3D editing workflows. Multiple motion paths can be combined and staged in the main camera track (\autoref{fig:basic panels}c1), resulting in composed sequences of camera and element movements. For example, the camera may move from a medium shot of a character’s hands fastening buttons to a close-up of the face, while the subject simultaneously steps forward to emphasize the slow walking motion.

\subsection{Image Styling and Motion}
The \textbf{Image Styling and Motion panel}, located on the right side of the interface, refines the visual appearance of the scene in each camera shot. A selected camera view is displayed to the user, and the user has as slider (\autoref{fig:basic panels}d1) which can be used to toggle between the raw 3D image input and the restyled output continuously. The style of the shot, and high-level motion of the elements can be specified through text within the interface.

\begin{figure*}[ht]
    \centering
    \includegraphics[width=\linewidth]{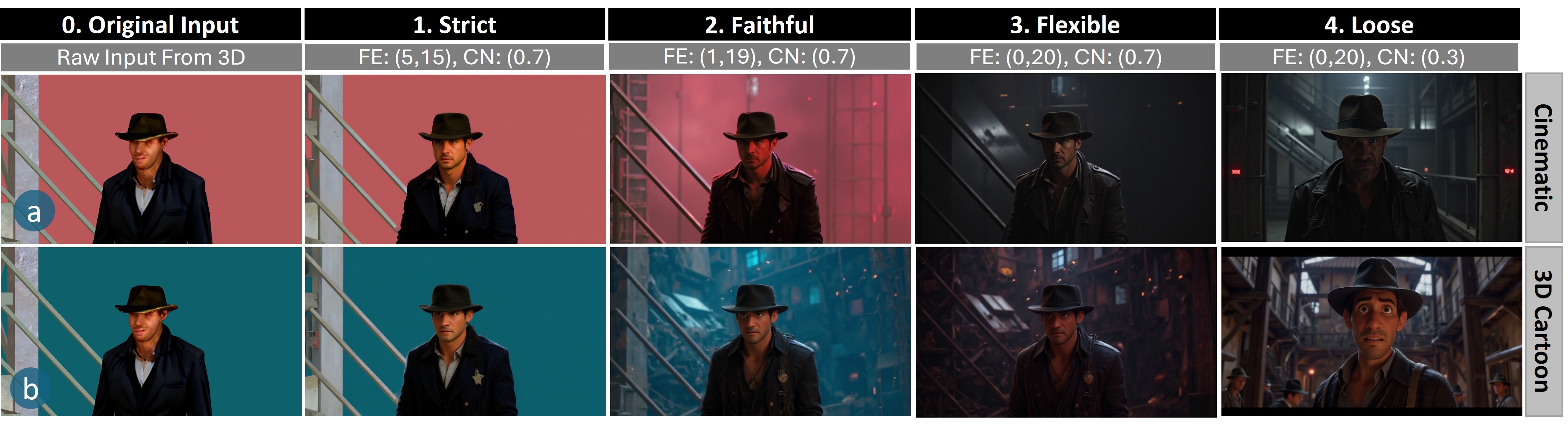}
    \caption{\red{
   Example adherence to source color in image generation with visual styles enabled by LoRAs: (a) Cinematic; (b) 3D Cartoon. Each row shows four resemblance levels (Strict, Faithful, Flexible, and Loose) which progressively relax the degree of adherence to the original 3D input (column 0). The FE values shown above each column refer to FlowEdit parameters, and the CN values refer to ControlNet parameters, which together control how strongly the generated images follow the source 3D frame.
    }}
    
    \label{fig:resemble-vs-decouple}
\end{figure*}

\begin{figure*}[ht]
    \centering
    \includegraphics[width=\linewidth]{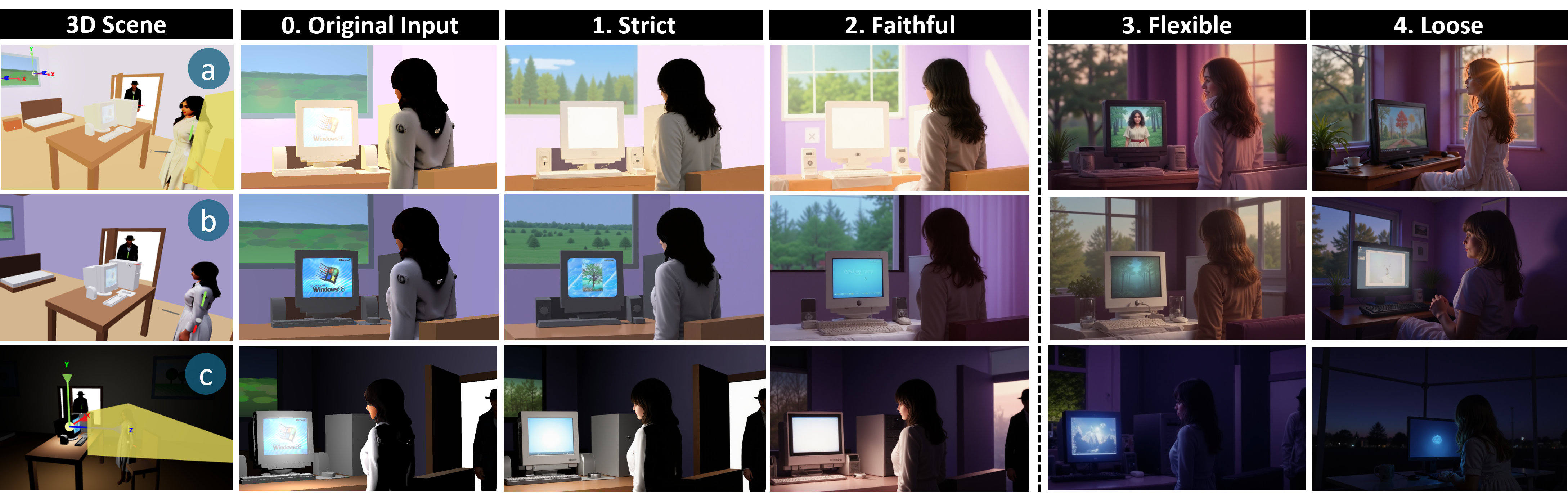}
    \caption{\red{
  Scene with Lighting variations (a) Sunny Day (b)  Dawn/Early Morning (c) Dark Room with four resemblance levels.
    }}
    \label{fig:levels of adherence}
\end{figure*}




\paragraph{Genres and Styles}
Users can define the target output by specifying both a genre and a style, along with the desired mood and tone. Genres (e.g., romance, dystopian, horror) primarily shape the story themes and narrative patterns, whereas styles (e.g., anime, film noir, documentary) determine the aesthetic and visual approach used to present the story. The style can have a dramatic effect on the scene, even with the same 3D composition (\autoref{fig:levels of adherence}), a style of Cinematic (a) will look very different from Cartoon (b).


\begin{figure*}[ht]
    \centering
    \includegraphics[width=\linewidth]{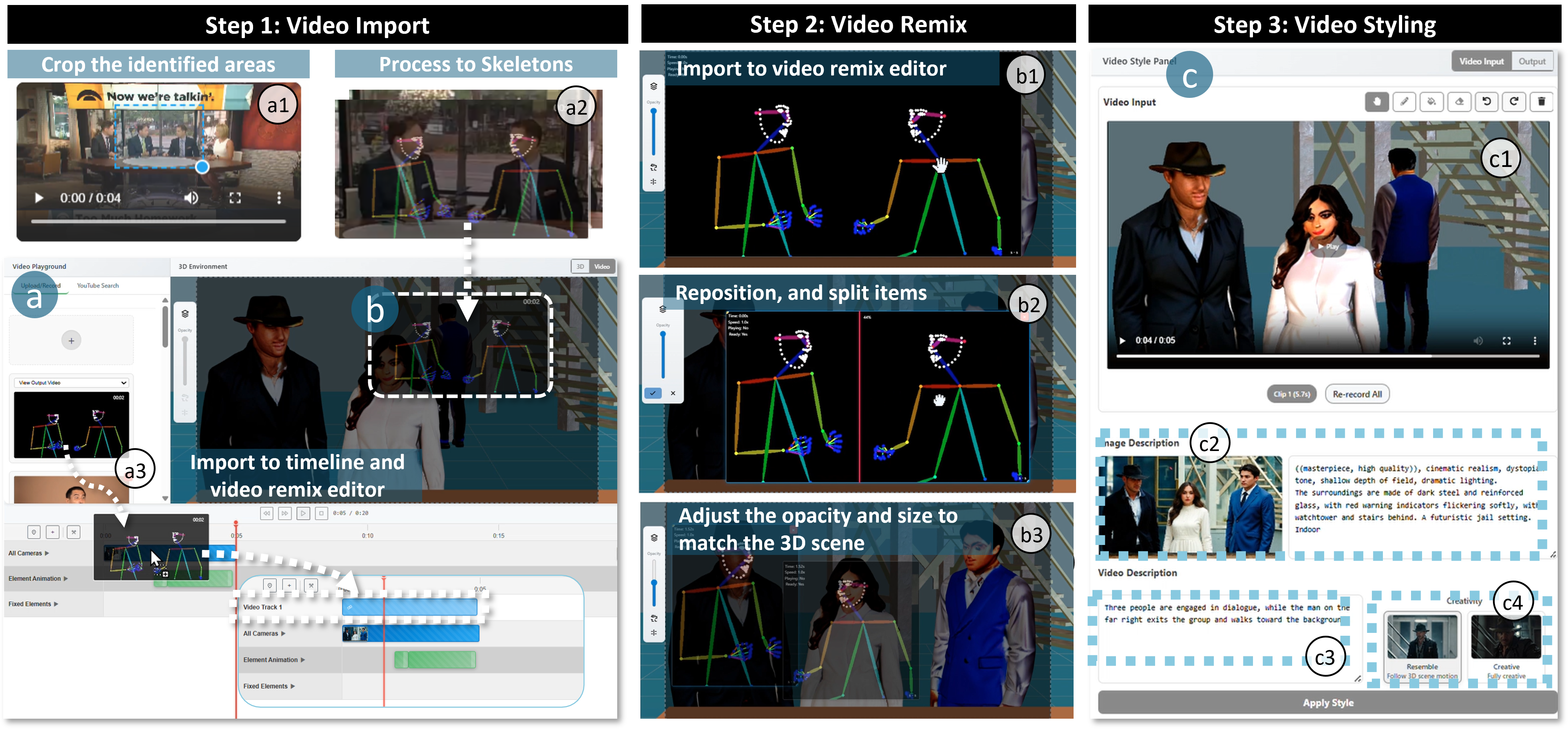}
\caption{\red{Demonstrating athe Video Style Interface with a three-person interaction example: (a) Video Import Panel: Users can (a1) import online or live video footage,  crop it, and (a2) process it into skeleton videos. These skeletons can then be (a3) dragged onto the timeline, where (a5) a new video track is created to guide character movements in the scene.
(b) Video Remix Editor: provides tools for manipulating, aligning, and refining processed video layers with character positions in the 3D scene (e.g., orientation, gestures). Users can (b1) resize and reposition clips, (b2) split them, and (b3) arrange the split segments to match character scale before recompositing them into a guiding video.
(c) Video Style Panel: includes (c1) the processed external video inputs; (c2) image descriptions and references from the image style panel; (c3) a video description field where users can specify additional movement details; and (c4) a resemblance–creativity control, where Resemble follows spatially defined motion, while Creative generates outputs based on the text prompt. }}
    \label{fig:layers}
\end{figure*}

\paragraph{Controllable Adherence to Colour and Lighting}
Once genre and style are defined, users can choose to (i) preserve the original (rough) render, (ii) use prompt-driven restyling (\autoref{fig:basic panels}d3), or (iii) only loosely follow the spatial composition and allow for more creative generations.

\systemname provides four levels of resemblance (\autoref{fig:resemble-vs-decouple}) \red{by utilizing the FlowEdit method that maps between the source (3D environment) and target distributions (output image). 
The underlying design intent is to mirror to what extent filmmakers evolve a shot from rough layouts to polished aesthetics.}
\begin{itemize}[leftmargin=*]
\item \textbf{Strict}: firmly preserves the original composition and colour palette.
\item \textbf{Faithful}: retains much of the original colour scheme while allowing moderate creative variation.
\item \textbf{Flexible}: diverges from the original colour and lighting, guided primarily by the text prompt, but maintains the spatial composition.
\item \textbf{Loose}: departs from both the original colour style and spatial composition.
\end{itemize}

\red{The structure of these four levels is informed by both iterative experimentation and insights from prior work on controllable generative imaging (\eg, ControlNet-like spatial conditioning and latent blending~\cite{croitoru2023diffusion,kulikov2024flowedit,zhang2023adding}), but are presented to users as intuitive artistic controls.}
For instance, changing a backdrop colour (\autoref{fig:resemble-vs-decouple}a/b) produces outputs that approximate the desired tone. In contrast, `Loose' uses ControlNet~\cite{zhang2023adding} alone, which applies depth-guided conditioning without latent blending, enabling full prompt-driven modifications such as cinematic lighting or dystopian tones (\autoref{fig:resemble-vs-decouple}Loose).

This choice determines how closely the output reflects the colour and lighting of the raw 3D input. Additional constraints can also be applied directly within the 3D scene, such as modifying base asset colours or experimenting with lighting such as balanced versus contrasting lighting (\autoref{fig:levels of adherence}).


\subsubsection{Video Style Panel: Restyling Simple 3D Motions}
Toggling from the 3D view to the video tab (\autoref{fig:basic panels}b3) transforms the interface from a 3D Editor into a Video Preview, with all sub-tracks of cameras and elements hidden, and shown as a video track, after the animations are exported. This shift signals that the rough spatial and scene blocking has been fixed, and the focus now moves to restyling the recorded videos.

On the right side, the Image Style Panel will transition into the \textbf{Video Style Panel} (\autoref{fig:layers}c), supporting video-specific prompt input. 

Playback of the timeline will show the 3D animation previews in the Video Output Panel (\autoref{fig:layers}c). However, when the user selects a specific video clip, the system highlights it in the panel as a dedicated preview, which can be played back independently in video format (\autoref{fig:layers}c1). This supports fluid navigation between reviewing the full timeline and focusing on individual clips for restyling.

If a styled image is attached to the clip, both the image and its associated description are also shown beneath the video preview (\autoref{fig:layers}c2), providing a combined visual and textual reference for guiding restyling. If no image is attached, the area simply displays “No image yet” and remains blank.
The video text description (\autoref{fig:layers}c3) is displayed at the bottom of the interface, and users can extend or modify this description in the video prompt input field.

Furthermore, the resemblance-creativity control (\autoref{fig:layers}c4) allows the users to define whether the motion should follow the defined motion or purely based on the video description.
Once confirmed, the clip is submitted for video generation. During processing (approximately one minute), the clip status displays a progress bar, and the interface remains interactive—users can edit other clips, adjust styling prompts, or prepare additional camera and avatar movements.

When the processing completes, the generated video appears in the Video Style Panel and is synchronized with the corresponding clip in the timeline. Users can toggle between the rough input animation and the generated video output, enabling direct comparison and iterative refinement. 
In practice, clips are first defined through rough motion blocking and then restyled visually, before being exported in batch to the generative video model for final synthesis.

By combining rough 3D input with restyling controls, and allowing flexible adherence to the original setup, the Image Styling Panel provides a lightweight yet powerful workflow for exploring visual design choices during previsualization.

\subsection{Granular Motion Control}
After completing the scene setup and restyling, the user can assess the results, and add more granular motion control if needed. This may be needed if the movement generated in the styling phase does not match the intended motion closely enough, or specific movements, gaze directions, or hand gestures are desirable.

\subsubsection{Advanced Video Remixing with Video Playground and Video Remix Editor}
\label{sec:advanced}
Some more complicated scenes may include both spatial and temporal elements, such as a parallel action with the \textit{foreground} being a composition of three people speaking in front of a table, with a person moving from foreground to the \textit{background} to grab an object. While the user could block this out in the 3D scene, it might be more effective to record their own video to demonstrate the movement. This types of scene may need more fine-grained control rather than rough control such as text with more creativity.

\paragraph{Video Playground}
To go beyond basic blocking where motions are defined primarily from the time-based elements recorded from the 3D scene, users can import external videos (either online videos or self-captured video footage) as a reference in the Video Playground panel (\autoref{fig:layers}a). The movement of the characters in that video will be automatically extracted as 3D skeletons, and can be used to drive the generation of the resulting video.

The Video Playground (\autoref{fig:layers}a) shows the video library where users can import online videos or self-captured video footage into it, crop (\autoref{fig:layers}a2), and process (\autoref{fig:layers}a3-4) it into a processed movement (skeleton) video. The processed video can then be dropped into the timeline, which will appear as a video track (\autoref{fig:layers}a4-5). 
\begin{figure*}[ht]
    \centering
    \includegraphics[width=\linewidth]{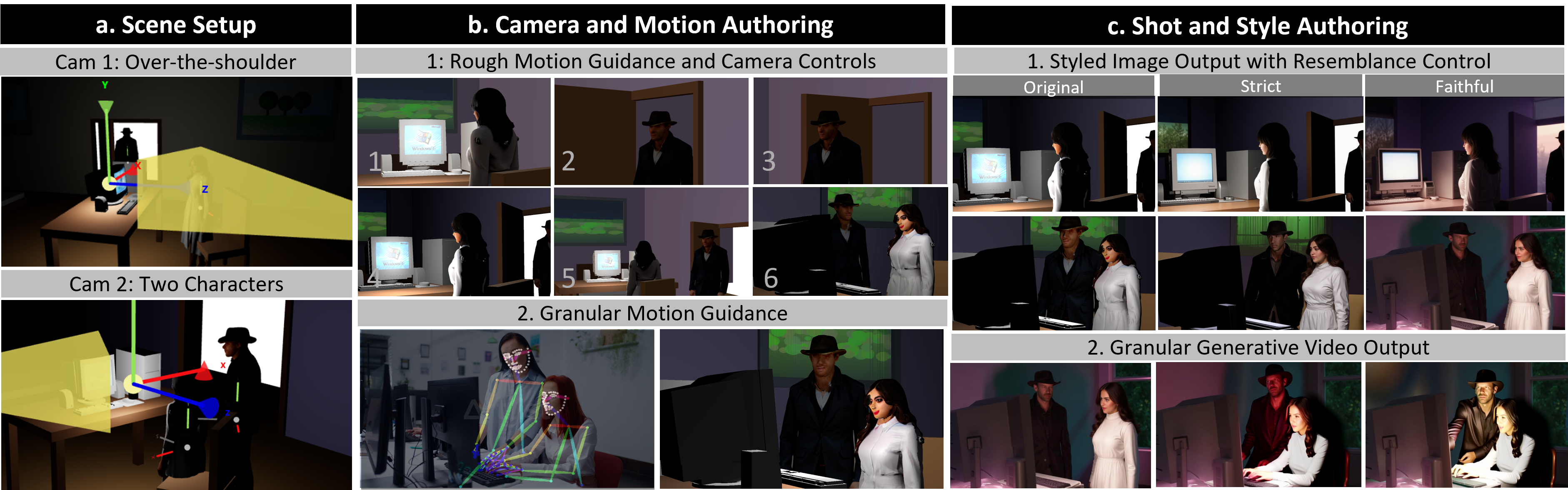}
    \caption{
    \red{\systemname Walkthrough Example:
(a) \textbf{Scene setup.} The director positions two characters in the 3D blocking panel, adjusts props, lighting, and colour, and configures complementary camera angles.
(b) \textbf{Camera and Motion Authoring.} The director first defines rough motion guidance and camera placement (1–6), then applies granular motion guidance using control-video references and skeleton alignment to refine gestures, body posture, and interaction timing.
(c) \textbf{Generated Shot and Style Authoring.} The Image Styling panel previews lighting presets with resemblance levels (Original, Strict, Faithful) to explore visual tone. Granular generative video output further blends the animated 3D blocking with external motion footage, enhancing realism, lighting continuity, and character interactions across the final sequence.}
    }
    \label{fig:walkthrough}
\end{figure*}

\paragraph{Video Remix Preview}
This processed video will then appear in the Video Remix Preview (\autoref{fig:layers}b), which will now serves as a video layer editor (\autoref{fig:layers}b1-3), where the user can resize and reposition the video (\autoref{fig:layers}b1), split it into multiple videos (to account for multiple people \autoref{fig:layers}b2), and resize these individual processed videos and reposition them to match the positioning in the composition defined in the 3D scene (\autoref{fig:layers}b3)
and then recomposite the split videos into one guidance video. 
This can also be imported in the video restyle panel (\autoref{fig:layers}c1) for the video generation process, similar to the output from the original camera output from the 3D scene.
Furthermore, this fine-grained motion (\eg, multiple people having a dialogue, \autoref{fig:layers}b3) can be remixed with the motion originally defined in the 3D scene (\eg, John moved to the back in \autoref{fig:layers}c1), as in \autoref{fig:layers}.
Blending the capability of 3D environment and 2D video sources, we can utilize the 3D environment to sketch out the spatial movements, but also provide more fine-grained movement composition, using 2D videos to control individual character's movement. 

While we illustrate the workflow using a small set of representative scenes and styles, we provide additional examples demonstrating generalization across visual styles, lighting conditions, scene types, and interaction complexity in the Appendix.

\subsection{\systemname Walkthrough}

Alice is a young director who is seeking to previsualize a short film that she is intending to pitch to secure funding to fully produce it. She would like high-quality renders that convey the narrative and tone of the film, but she does not have the time to learn complex 3D software, nor the budget to hire out the previsualization. She turns to \systemname to work through some ideas and generate the visuals for the first scene, in which a hacker is trying to gain access to a secure system when a conspirator approaches her to let her know time is running out. 

In this scene, the main character is working in a dimly lit room at a glowing computer \red{(\autoref{fig:walkthrough}a)}, so Alice uses the scene blocking panel and adds the hacker and conspirator characters to the scene, as well as the computer. She moves the characters and props into their initial positions. \red{She experiments with different lighting and color trials (\autoref{fig:levels of adherence}a-c) to establish the desired atmosphere, ultimately darkening the room and adding a subtle glow from the computer screen to create the intended visual mood (\autoref{fig:walkthrough}a)}. 
Alice adds two cameras to the scene - one to focus over the shoulder of the hacker onto the screen, and another that will frame both characters as they speak to each other. To add basic animation to her scene, Alice selects the conspirator and records a motion path, using the WASD keys to move the character along a trajectory towards the hacker. She then selects the over-the-shoulder camera, and adds a keyframe so that it follows the conspirator as he walks towards the hacker \red{(\autoref{fig:walkthrough}a-b1)}. 

With the basic blocking complete, Alice moves to the Image Styling and Motion panel (\autoref{fig:basic panels}d-g). She experiments with different restyling options, and gives the system more license to deviate from the 3D blocking by selecting `Faithful' adherence. She also specifies a rough motion direction of \textit{a man walks to a woman who is typing on a desktop computer, and speaks to her}. After iterating several times, Alice realizes that she likes the style, but the motion isn't believable.

Alice moves to the Granular Motion Control panel (\autoref{fig:layers}) to try combining the generated video with external footage to more accurately animate the characters. She finds an online video with a discussion scene, imports it into \systemname  \red{(\autoref{fig:walkthrough}b2)}. \systemname processes it into skeleton-based video guidance which she can then overlay onto the video to use as guidance. This processed video has gesture and facial expression details, allowing her not only to animate individual characters but also to capture the group dynamics between them (\autoref{fig:layers}b). Alice can also align the videos and mix the 3D motion paths of the conspirator (\autoref{fig:layers}b) with the 2D video guidance that shows the interactions between the hacker and conspirator to create more complex interaction dynamics.

After tweaking the alignment and generating a new video, Alice is satisfied with the final result \red{(e.g., \autoref{fig:walkthrough}c2)}. Through those three steps, Alice was able to block the scene in a way that she feels confident she will be able to film, and the resulting video clips are not only believable motions and expressions, but they convey the tone and mood that Alice is hoping to relay during her movie pitches. \red{Full key frame comparisons of the generated video output versus original 3D input are shown in Appendix (\autoref{fig:app-shot}).}

\subsection{Implementation}
The front-end web interface
was built using React+Vite and React Three Fiber. The back-end server was built using Node.js. The server implements a video processing pipeline that takes a video, crops and processes it using FFmpeg, and sends it to a ComfyUI back-end~\footnote{\url{https://github.com/comfyanonymous/ComfyUI}} for further processing. 
The ComfyUI workflows includes a set of JSON files that creates processed videos (depth, skeleton, bounding box, outline), composition workflows (for compositing videos such as depth, skeletons videos, and image-to-image workflows (with ControlNet~\cite{zhang2023adding}, Diffusion Models~\cite{croitoru2023diffusion}, and Flux~\cite{labs2025flux1kontextflowmatching,flux2024}, and Flow Edit~\cite{kulikov2024flowedit}), and image/text/video-to-video models (Wan 2.1~\cite{wan2025}, Wan Fun Control~\cite{wan2025wan}, VACE~\cite{jiang2025vace}).
For more effective authoring, the Prompt Composer (\autoref{fig:basic panels}f1-3) was implemented to help users formulate effective prompts without needing prompt-engineering expertise. It provided structured input fields for visual style, mood/tone, and description of background surroundings of the selected camera frame, which were expanded by calling a OpenAI GPT service\footnote{\url{https://openai.com/index/openai-api/}} into detailed natural language prompts. 
\begin{figure*}[ht]
    \centering
    \includegraphics[width=\linewidth]{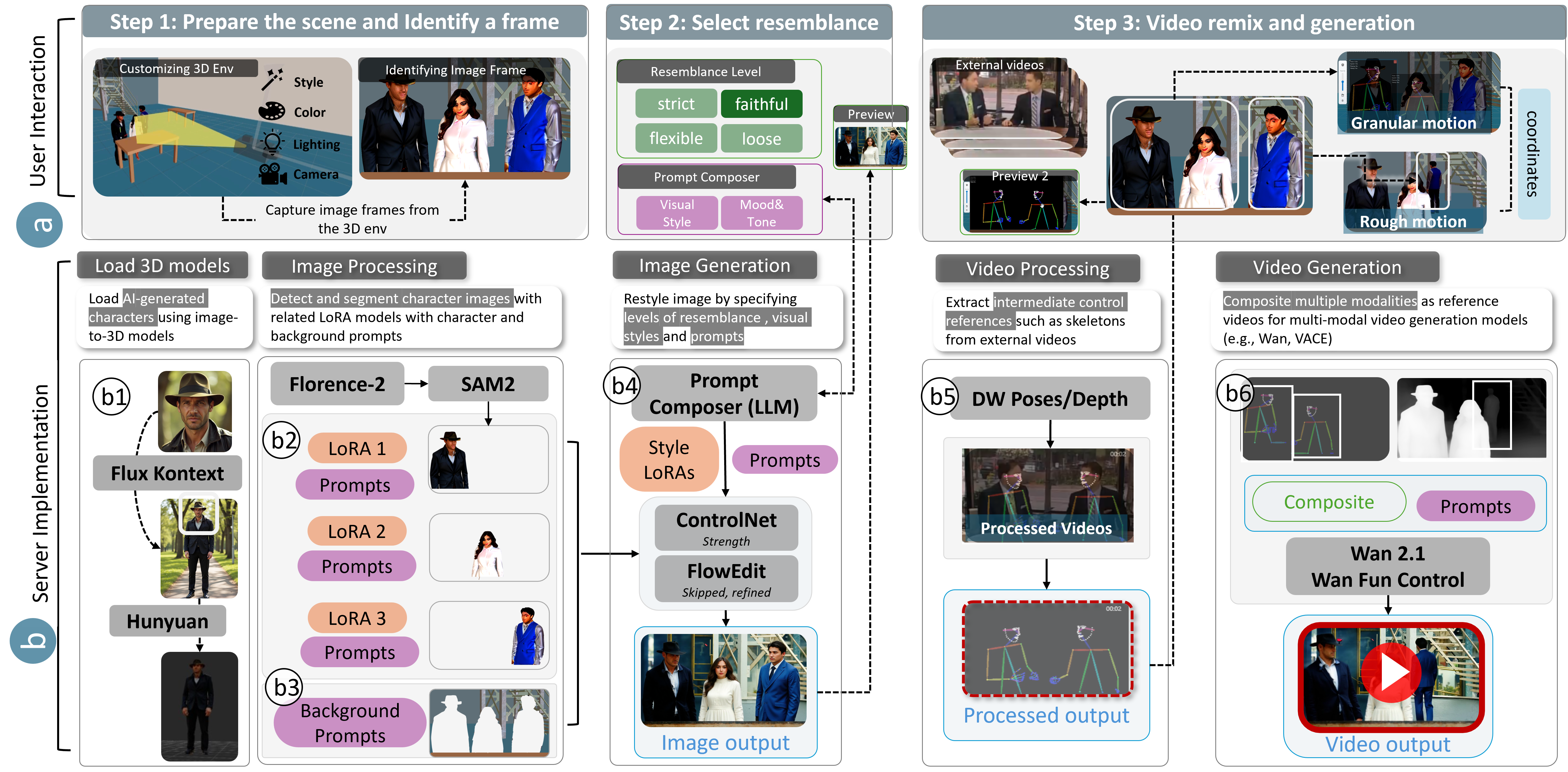}
\caption{\red{Overview of the PrevizWhiz Authoring and Generation Pipeline. The top row illustrates user-facing interactions: (1) preparing a 3D scene, (2) selecting resemblance levels, and (3) importing external videos for motion refinement. The bottom row shows the corresponding server-side implementation: (a) loading 3D models, (b) processing character regions with LoRAs, (c) generating restyled images, (d) extracting pose/depth references, and (e) composing multimodal inputs for video generation using Wan/Wan-Fun Control.}}
    \label{fig:implementation}
\end{figure*}
\subsubsection{Elements}
For the tutorial and sample scenes, the scene elements are composed of both procedural geometry and imported 3D models. Furniture (the table, chair, cabinet, and bed) is made of basic geometry such as boxes, styled with standard materials and appropriate colouring. More specific items (an old mirror, a desktop computer, and a watchtower) are GLB models downloaded from SketchFab\footnote{\url{https://sketchfab.com/feed}}.
\red{Furthermore, the 3D characters in our scenes were AI-generated (\autoref{fig:implementation}b1) using Hunyuan-3D~\cite{lai2025hunyuan3d25highfidelity3d} from 2D full-body character images. Because most of our reference images were half-body portraits, we first produced full-body versions of each character using Flux Kontext~\cite{labs2025flux1kontextflowmatching}, and then used these synthesized full-body images as inputs to Hunyuan-3D.} 

\subsubsection{Character and Style Low-Rank Adaptation (LoRAs)}
To augment the capabilities of massive pre-trained models like diffusion models, LoRAs~\cite{hu2022lora} are fine-tuned training of an existing model with new weights that aimed to enable more specialized knowledge or style, which enables character consistency and different stylizations within \systemname.

\textit{Character LoRAs.}
Starting (for instance) from an image of 3 characters captured from the 3D scene, the system resized that camera image, grounded three “person” instances with Florence-2~\cite{xiao2024florence} using bounding box detection for each detected "person". 
Each Florence box is sent to SAM2~\cite{kirillov2023segment} to turn coarse boxes into pixel-accurate masks per character; their masks were softly expanded (expand ~12–18 px, blur ~3.5–6) and composited to separate each character from the background image, so edges do not leave seams when restyled. All the character masks are then composited together, and the result is inverted to get a clean background mask.

\red{Each region of segmented character (\autoref{fig:implementation}:b2)} then received its own conditioning: \textit{character-specific prompts} with identity LoRAs were applied via CLIP hooks, while the \red{\textit{background prompt} (\autoref{fig:implementation}:b3)} was guided separately (e.g., a cinematic, dystopian jail of steel and glass with subtle red warnings). By separating characters from the background image,
this yields a single composite conditioning: characters are steered by their own prompts/LoRAs inside their masks, while the environment is driven by the background prompt elsewhere.

\textit{Style LoRAs}
Guiding styles with purely text description can be inconsistent. Therefore, the visual styles that were chosen in the image style panel were also guided by style LoRAs with Anime, Cartoon, Pixel Art, and Realism. As LoRAs are trained and released, they could be added to the system to allow for even more flexibility in the future.

\subsubsection{FlowEdit and ControlNet Parameters}
\red{The generation process (\autoref{fig:implementation}:b4) for the resultant images was run for a total of 20 steps. Four resemblance levels were created by adjusting skipped steps and ControlNet strength during a 20-step generation process:} (1) Strict: Skips 5 steps with a ControlNet strength of 0.7. It strongly preserves the original image's colour, style, and spatial structure; (2) Faithful: Skips 1 step with a ControlNet strength of 0.7. This level also preserves the original's colour, style, and structure, but slightly less than Strict.
(3) Flexible: Skips 0 steps with a ControlNet strength of 0.7. It maintains the original spatial composition but generates new colours and styles.
(4) Loose: Skips 0 steps with a reduced ControlNet strength of 0.3. It diverges from the original in colour, style, and spatial structure, offering the most creative freedom. 
\red{The four resemblance levels were chosen based on the authors’ experimentation for intuitive artistic control and are also informed by prior work~\cite{zhang2023adding, kulikov2024flowedit}.} These were representative for showing the difference between the four levels of adherence, but can be varied when the user choose different number of steps, and is also dependent on the roughness of the original 3D scaffolds. 



\subsubsection{\red{Skeleton Extraction and Recomposition Pipeline For Video Generation}}
\red{To extract motion from external videos (\autoref{fig:implementation}:b5), PrevizWhiz processes each clip by generating  skeleton keypoints and depth maps for every frame. Users can crop, reposition, and rescale the processed skeleton video in the Video Remix Editor so that character positions align with the 3D scene’s spatial layout. After adjustment, the system recomposites the skeleton and depth sequences (\autoref{fig:implementation}:b6) into conditioning videos, which are then used as multimodal inputs to the generative video model. This allows the final output to follow both the 3D blocking and the fine-grained gestures derived from the reference footage.}
\subsubsection{Hardware Configuration}
\label{sec: imp-setup}
The generative image and video model was run on a separate RTX 5090 VRAM24GB, 64RAM computer, and accessed via local network via http post. 
We used 1.3B VACE and WAN models for the user study as it is a more performant model and allows for quicker generation of images.
\section{Evaluation}

\begin{figure*}[ht]
    \centering
    \includegraphics[width=\linewidth]{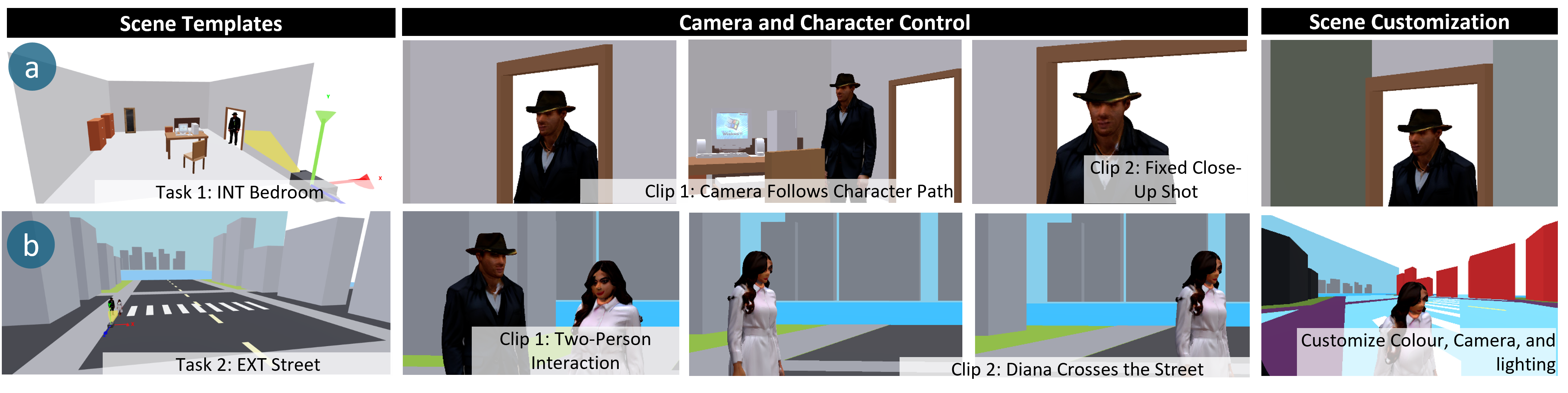}
    \caption{Study Scenes and Tasks (a) Task 1 with INT Bedroom; (b) Task 2 with EXT Street}
    \label{fig:study-setup}
\end{figure*}

\begin{figure*}[ht]
    \centering
    \includegraphics[width=\linewidth]{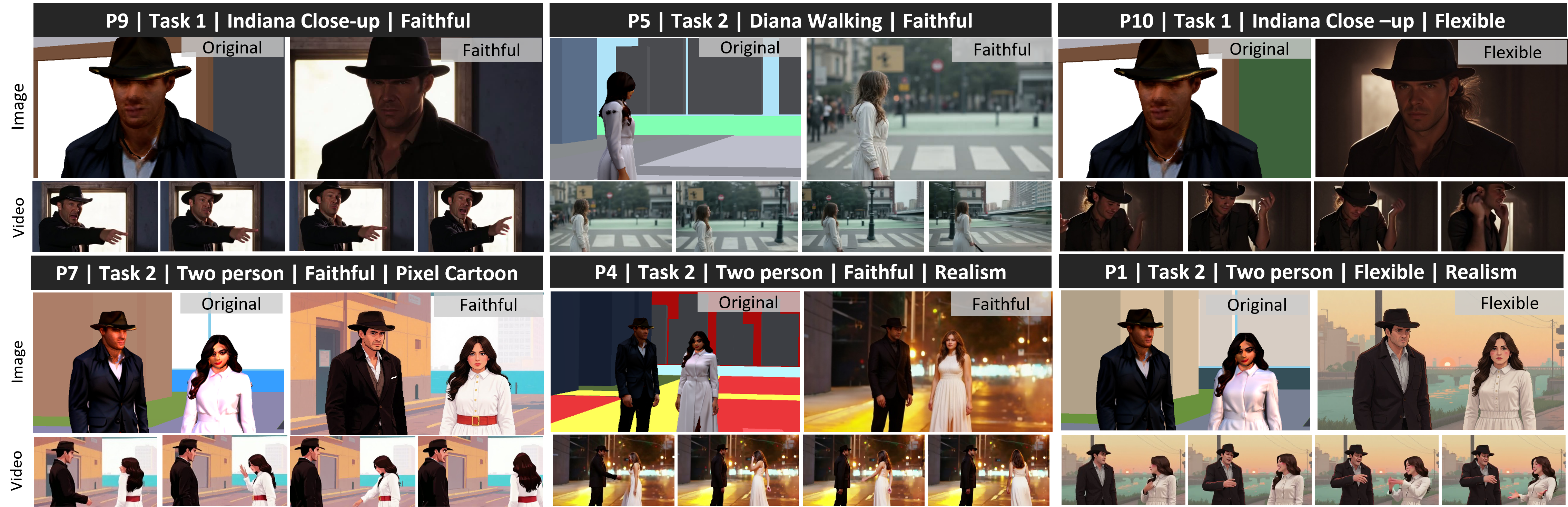}
    \caption{Images and Videos Output from the User Study where users tried different text prompts, visual styles, guidance video, and change of color, lighting, camera position and lens to achieve the final video output.}
    \label{fig:study-output}
\end{figure*}

\begin{figure*}[ht]
    \centering
    \includegraphics[width=0.8\linewidth]{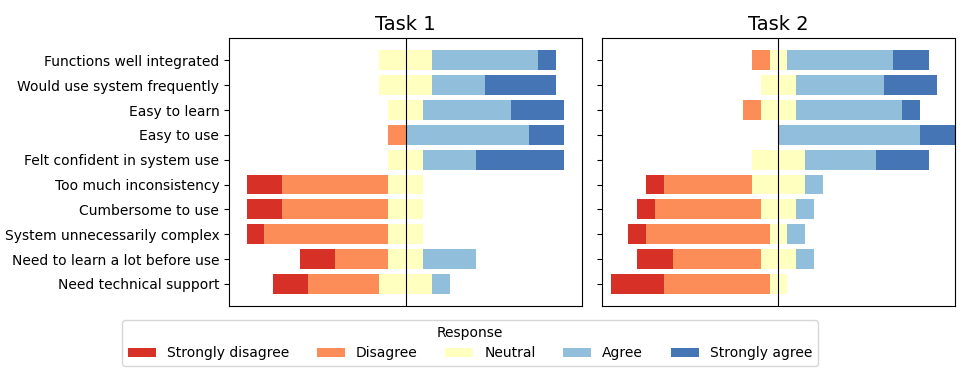}
    \caption{An Overview of the Responses for the System Usability Scale Questionnaire}
    \label{fig:data}
\end{figure*}

We conducted a study to understand how filmmakers would use \systemname to generate previsualization content. 
 The study was designed to gather both quantitative and qualitative observations and feedback from filmmakers as they used the different components and functionality of the system. The study was reviewed and approved through our institutional ethics review process.

\subsection{Participants}
We recruited 10 participants (5 female, 4 male, 1 non-binary), aged 23–42, through targeted outreach, community channels, and snowball sampling. Eight were filmmakers and creative professionals (cinematographers, directors, advertising directors, technical directors, and costume/styling specialists, three of them also having experience with 3D tools), while the remaining two were 3D/animation experts who have worked in the filmmaking industry.

Participants had 1–15 years of experience ($Mean = 7, SD = 4.3$), spanning early-career to highly experienced professionals (full details in Appendix~\autoref{tab:demographics}). Participants background and use of related technologies varied. Six reported active use of 3D pipelines, while the others had limited 3D exposure. Seven participants had used image generation tools (\eg, MidJourney, ComfyUI) and five had explored video generation (\eg, Runway, Veo, Kling), often for storyboards, pitch decks, posters, or exploratory sequences. Six participants used 3D previz tools (Previs Pro, Cine Tracer, Set a Light 3D), three relied on traditional methods (storyboards, animatics, motion capture), and one reported no previz use. Their production work ranged from large-scale features and advertising, which relied on team collaboration, to independent films and small-scale content creation, reflecting a spectrum from high-budget studio pipelines to resourceful indie practices.
Participants received \$135 USD and the study took 90-120 minutes. 

\subsection{Procedure and Tasks}
After completing the consent form and demographics survey, the participants watched a tutorial explaining the system and how to use it, followed by two study tasks where they used the system to generate previsualization renders for an interior and exterior scene. Following the tasks, participants completed a survey and took part in a semi-structured interview. The experiment was conducted using the same hardware apparatus described in \S\ref{sec: imp-setup}. 

\subsubsection{Tutorial Videos (10-15 min)}
Participants watched a tutorial video about the system and the experimenter explained the features and functionality of the system. The tutorial scene was of a person walking, and was different from the scenes used in the two study tasks. Throughout the study, the experimenter was also available to answer questions if the participants needed assistance as they completed their tasks.

\subsubsection{Study Tasks (50-60 min)}
We provided two tasks with two rough scenes where users are required to change the background colour or lighting to iterate on and create their own scene. There are pre-set cameras added to the scene that participants can change the positions, angles, and FoVs to create their own narrative. 

\paragraph{Task 1 (20min) Single-Person Scene}
The experimenter described the interaction around features like camera blocking, character movements, image and video generation process, and asked the participant to perform the specific interactions.  
The user was given a script for a bedroom scene (\autoref{fig:study-setup}a):
\begin{quote}
\textbf{INT. BEDROOM}

    Indiana stands by his bedroom door, and walks  into his bedroom.
\end{quote}

Given this script, the participant was asked to create two continuous video clips utilizing two pre-set cameras. One camera will need to be moveable for capturing the character's motion path (walking into the bedroom) and another camera will be a fixed close up view capturing the person's facial expression, guided by a set of pre-loaded external video clips. Participants were also required to change the colour and lighting of the walls.


\paragraph{Task 2 (25min) Multi-Person Scene}
Similar to the previous task, the experimenter introduced the features through a multi-person dialogue scene. This task emphasized additional functions such as detailed LoRA-based character descriptions and video layer splitting. 
The example and scene used for this task is a street view (\autoref{fig:study-setup}b) and the multi-person dialogue with the following script:
\begin{quote}
\textbf{EXT. STREET}

Diana chats with Indiana on the sidewalk, then Diana leaves to cross the street.
\end{quote}
In this task, participants were required to create two clips, where in the first clip there will be two characters speaking, and in the second clip, one of the characters leaves and walks across the street. \red{Although key scene elements were fixed for comparability across sessions, the task was designed to provide substantial creative freedom.} Participants could modify the colour, lighting, and style of the street scene (including the street, sidewalk, ground, and surrounding buildings) and define the camera shots and movements. 
This task allows us to examine multi-character shoots with external video sources being used to animate more complex interactions, where the participant will need to edit and remix the guidance videos. 

\subsection{Data Collection and Analysis}
After each task, participants completed the System Usability Survey (SUS). Once all tasks were finished, participants filled out a post-study survey probing system features and user experiences, followed by a semi-structured interview with in-depth questions about their experience, impressions of the system, the authoring levels we provided, and potential applications. All the text, image and video input and outputs were logged, and the session was video and audio recorded.

\red{We analyzed the qualitative data using inductive thematic analysis~\cite{braun2006using}. We first applied open coding to participants’ verbalizations and interview responses to identify recurring concepts and usage patterns in the system. The codes were iteratively refined and grouped into higher-level categories, and emerging interpretations were regularly discussed among the co-authors to ensure consistency and analytic rigor.
This bottom-up process ensured that themes emerged directly from participant experience.}


\section{Findings}
In our analysis we group insights into three main themes: (1) flexible and effective authoring workflow, (2) balancing control and creative exploration, (3) perceived benefits and concerns of using AI in pre-production.
Below we summarize each theme.


\subsection{Flexible and effective authoring workflow}
Overall, participants found \systemname as easy to learn, fast to operate, and well-suited to quick iteration, highlighting straightforward camera/character blocking in 3D and rapid preview generation. \autoref{fig:data} shows the SUS survey results.
Responses to the post-study survey indicate that recording camera and character motion paths to guide video generation was helpful ($Median=4, IQR=1$ for camera; $Md=4, IQR=0.5$ for characters), that style control felt flexible ($Md=4, IQR=1$) and that users experienced solid creative control ($Md=4, IQR=0$). They further reported that the generated outputs made sense to them ($Md=4, IQR=1$). 
At the same time there was mixed agreement on whether the generate content matched their imagination ($Md=3, IQR=0.5$), underscoring a tension between rapid output and alignment 

Compared to familiar tools, participants perceived speed advantages over traditional approaches ($Md=5, IQR=2$) and comparable output quality from the generative pipeline ($Md=4, IQR=0.5$).
Several participants contrasted \systemname with the steeper learning curves of professional previz software (e.g., C4D, Cinetracker), noting that it required less training yet produced usable outputs; one participant summarized it as \textit{``basically like an AI version of an editing software [...] that makes things more convenient''}, reducing the need to juggle multiple tools like MidJourney and Runway (P9).

\subsubsection{Integrated Workflow of 3D-to-2D as a Middle Ground Between Precision and Flexibility}
Participants saw the 3D-to-2D workflow as a practical middle ground: faster and cheaper than building full high-fidelity 3D pipelines (\eg, Cine Tracer), yet offering more structure and control than text-to-video or image-to-video approaches (\eg, Midjourney, Runway). This positioned \systemname as both efficient and reliable, especially for early creative alignment.

Several participants noted that this balance reduced the risk of costly revisions by clarifying ideas upfront. For example P9 explained \textit{``Its form of presentation also helps reduce the number of revisions we need to make and lowers the overall communication cost.''} 
They further highlighted benefits in pitching contexts, where client expectations can diverge from professional assumptions: \textit{``If the client isn’t from this professional background, what they imagine might be completely different. [...] Using this kind of software [...] doesn’t require significant financial investment. [...] Even saving around 10\% makes this approach worthwhile.''}

These findings suggest that \systemname not only accelerates authoring but also serves as a communication bridge between filmmakers and stakeholders with varying levels of technical expertise. 
\subsubsection{Rapid Iteration and Experiments with Restyling}
Participants valued the ability to move quickly from rough scaffolds to polished outputs, using restyling to test diverse aesthetics such as cartoon, realism, or cinematic (see \autoref{fig:study-output} for examples from the user study). This workflow enabled them to generate multiple options in a short time and experiment fluidly with color, lighting, and camera placement.

Participants described this process as both accessible and productive. P6 emphasized that it was \textit{``very accessible, easy to use, and able to generate a multitude of options in a relatively short time,''} while P8 noted the benefit of being able to \textit{“roughly, quickly, get to an idea... and then very quickly experiment with different final aesthetics.”}

Overall, these findings highlight the value of a rough-to-polished flow: starting from low-fidelity scaffolds to quickly explore composition and then refining towards higher-quality outputs. \systemname functioned not only as a previz tool but also as a creative sketching tool, supporting rapid experimentation without locking users into a rigid production pipeline.

\subsection{Balancing control and creative exploration}
Participants emphasized that \systemname supported creativity across multiple dimensions (color, lighting, composition, motion, and costume) while also providing different degrees of controllability depending on their professional role and creative goals.


\subsubsection{Role-Specific Creative Priorities}
Different participants prioritized different aspects of the workflow. For example, P5 (working in styling and costumes), valued early experimentation with lighting and props to ensure costume palettes matched their environment and actors. 
In contrast, P7 (working in animation) highlighted the importance of refining fine-grained character movement, which went beyond what could be achieved with keyframes or stop-motion approximations. These differences underscore the tool's cross-disciplinary value.


\subsubsection{Balancing Controllability and Creativity}
Most participants preferred outputs that preserved core elements of the scaffold while still leaving space for stylistic variation. Seven of ten favoured the \textit{Faithful} resemblance setting, appreciating its balance between adherence and polish: P5 noted, \textit{``The 'Faithful' feels very controllable and has a sufficient expected style''} (P5). And P6 noted that “\textit{It maintains control over all of the parameters [...] but keeps the possibilities open for surprises.''} 

Three participants leaned toward more open-ended modes. P3 found \textit{Flexible} useful for \textit{``enhanced light and shadow for greater mood,''} while P8 valued how it left \textit{``room for my text prompt to actually have an aesthetic impact on the final result.''} P9 preferred \textit{Loose}, describing it as \textit{``more realistic, with richer details, more cinematic qualities, and more unexpected results.''}

\subsubsection{Preferences for resemblance vs. creativity in video outputs.}
A similar pattern appeared in video generation. Seven participants preferred \textit{Resemble}, citing its predictability (\textit{``The movement basically satisfied user’s requirement, nothing unnecessary''} - P2), while three opted for \textit{Creative}, describing it as more expressive and thought-provoking (P3: \textit{``more fun and thought-provoking''}). Some noted that both modes were situationally useful: Creative could introduce unexpected tension or dynamism, while Resemble ensured fidelity to planned scaffolds (e.g., P3, P9).



\subsubsection{Motion Scaffolds: Coarse-Grained vs. Fine-Grained Movements}
Participants highlighted the value of being able to work with different motion fidelity  levels depending on their goals. For early blocking, coarse trajectories, such as walking paths or camera pans, were often ``good enough'' to test pacing and shot composition. Several participants noted that this level of control felt fast and lightweight, lowering the barrier to experimentation.

At the same time, participants appreciated being able to refine toward fine-grained motion when needed. Detailed gestures, gaze, or expressions were described as critical for certain storytelling moments and having options beyond rough blocking allowed them to capture subtle movements that go beyond what can be done in traditional previz tools using keyframes or stop-motion animations.

\subsubsection{Misalignments of Cross-Modality Considerations}
Participants also noted instances where different modalities failed to align in conveying emotion or intent. For example, P7 noted that animations often rely on exaggerated movements to communicate purpose and drama: \textit{``They over-exaggerate movements to make it look purposeful...there are even nonsense frames, extreme movements, that make the action feel more intentional.''} This reveals a gap between realistic capture and stylized expressiveness that animation professionals often rely on. 

Another challenge was the need to manually reconcile text-based descriptions with visual elements in the 3D environment. P5 suggested the system could be improved by automatically detecting mismatches, such as when a prompt specifies a somber tone but the scaffolded scene retains bright, high-contrast lighting.




Taken together, these findings suggest that \systemname enables flexibility across levels of motion fidelity supporting both broad-stroke exploration and micro-level expressive details. It allows filmmakers to choose their position on a spectrum of control and creativity--from precise control to open-ended generative variation--depending on the task, role, and stage of pre-production. At the same time, occasional misalignments across modalities highlighted the need for strong feedback and integration

\subsection{Benefits and concerns of AI in pre-production}
Participants expressed mixed views on the role of AI-generated video in previz and production. While most saw AI as an \textit{auxiliary tool} that supports existing workflows, some seeing broader use in smaller production teams with limited resources. At the same time, participants also raised concerns about controllability, professional norms, and collaboration dynamics.


\subsubsection{Professional polish and external communication.}
Participants consistently reported that AI-restyled outputs helped them present ideas more professionally to non-expert stakeholders (\eg, investors, grant reviewers), reducing reliance on stakeholders' imagination.
This contrasted with rough 3D or screen-grabbed outputs from conventional tools. 
For example, P8 stated polished AI previews \textit{``would probably make me look more professional,''} and would likely perform better in grant contexts than stills from traditional previz tools \textit{``because it requires a lot of imagination [...] to look at 3D dolls''}.
P3 similarly differentiated audiences: rougher previews suffice for internal teams, but external audiences tend to need more refinement.

\subsubsection{Adapting for varied teams and workflows.}
Participants viewed the system as broadly applicable across production styles and scale, and useful for \textit{``simple videos or films``} (P3), while acknowledging that fit depends on team and director preferences. 
Directors’ working styles strongly influenced how previews were used. Some preferred strict alignment, demanding \textit{``demand[ing] that the footage be identical to what was filmed. [...] Others are more casual [...] That reflects the director's style.'' (P3)}  

Some contrasted bespoke setups with pre-made scene libraries, noting AI-assisted previs can reduce setup overhead: \textit{``For example, if we want two people fighting on a Hong Kong–style street, I just pick from hundreds of scenes that 3D staff provided, load one, and then place the characters. But then I have to adapt my shot design to whatever the scene already has''} (P9). With AI assistance, shots might not be picked from a library, but can be customized more easily.

These observations suggest that while \systemname can be adapted across workflows, its role depends heavily on team structure and the director’s creative approach, and to what extent they will adopt AI.

\subsubsection{Controllability and alignment with intent.}
A recurring concern was whether AI outputs can reliably match scripts and creative intent. Several participants described current pipelines as insufficiently controllable, positioning AI use as exploratory or partial rather than end-to-end. For example, P9 noted that \textit{``The downside [...] is that it's not very controllable [...] still takes time to refine. [...] We don't dare promise clients [...], only that we might use it to \underline{assist}. [...] if it works for the whole thing, great. But if not, then it just serves as a reference.''}
Relatedly, participants noted cross-modal misalignments (\eg movement exaggeration vs. intended emotion; tonal prompts vs. lighting), underscoring the need for stronger feedback and integration across text, style, and motion.

\subsubsection{AI and creative labour in filmmaking.}
\label{sec:ai-creative-filmmaking}
A recurring theme in discussions of AI in filmmaking is the fear of displacement: participants noted that while generative systems can accelerate previs and creative ideation, they also raise concerns about how such tools might reshape budgets, roles, and job security. P9 worried about how client perceptions of AI could directly translate into reduced resources: \textit{``Another reason we don’t rely too much on AI for previz is budget-related. If clients know that AI can be used, they might cut the budget. For example, originally a film might have a budget of x, but they might think, 'Since you can do this with just one or two people using AI, could you do it for y?' ''} P3, who worked within larger production teams, voiced similar concerns. She highlighted that full reliance on AI-generated content (AIGC) risks undermining established team structures, turning collaborative craft into a cost-saving measure: \textit{``We still tend to have a team-based division of labor. There's no way to fully utilize AIGC. If we did, it would only mean cost savings and eliminating everything else.''} She further stressed the implications of fully AI-driven pipelines: \textit{``One person can do it... That would leave us unemployed. We were supposed to be doing this before.''} 

These perspectives reflect broader industry anxieties: while AI may speed up certain tasks, it also threatens to collapse professional roles into smaller, less specialized teams, particularly in studio-scale environments where collaboration is the norm.

Yet not all participants viewed this shift negatively. P6, working in smaller-scale productions of short drama films, offered a more optimistic view based on his experience with AI-generated short dramas. He described a tiered perspective on AI's role in production:
\textit{``The top priority is pure AIGC, meaning everything is fully handed over to AIGC for direct generation, no live shooting at all. The next best approach is generating in 3D and then doing transfer, which is similar to the workflow your software supports now. And only after that comes the workflow of starting with live-action shooting, then turning it into 3D, and then into AIGC transfer. Because each step increases in difficulty and complexity, but at the same time, the ceiling of what you can achieve also rises. So, the core principle is: if AI alone can't handle something, or if simple tools can't solve it, then we resort to 3D scenes and real live-action shooting to tackle the more complex cases and get them done.''} 
For P6 AI-driven filmmaking was not a threat, but an opportunity: it lowered barriers for independent creators and small teams, enabling them to produce work that would otherwise be impossible given their constraints. With \systemname he especially valued the use of rough 3D scenes as a bridge between lightweight AI outputs and more controlled, polished results--suggesting a hybrid path where AI augments rather than eliminates creative labour.

\section{Discussion}
The user study revealed insights on how filmmakers interacted with the 3D scene and generative video, and opened up new questions around the utility of AI and the fidelity of generated media.

\subsection{Scene Blocking}

Participants appreciated being able to define colours, lighting, and camera angles within the 3D environment, which allowed them to establish the overall composition, tone, and feel of the shot. They also valued the flexibility of combining motion paths authored directly in 3D with fine-grained movements derived from external 2D video inputs, which is aligned with \textbf{R1} in enabling a flexible scene setup. 
We found that directors appreciated the simplicity of creating this rough scene that can be then restyled into polished video output, but would like to play more with the image and video generation side, where they typically instruct other staff to perform the more technical operations.  

Participants appreciated that \systemname did not rely on access to high-fidelity movie datasets~\cite{wei2025cinevision} or detailed 3D environments~\cite{rao2023dynamic}, and that they could use existing characters and objects and widely available videos to create their own scenes. This suggests future work of including some assets AI from existing scenes, or detexture some existing scenes as starting point so that users can use that to define rough color and lighting to craft their own scenes.

\subsection{Creativity Through Multi-Modal Intent}
Using \systemname as a prototype, we examined how users engage with rough 3D scene setup, stylized previews, and motion generation through generative video. The workflow enabled them to restyle environments at varying levels of resemblance, balancing fidelity to the original scene with creative reinterpretation guided by text prompts (\textbf{R2}).
Participants highlighted the advantages of combining motion paths defined in the 3D environment with reference movements from external 2D videos, which supported a balance between flexibility and expressiveness, which aligns with earlier design rationale (\textbf{R3}) aimed at supporting multiple levels of fidelity.
Participants generally preferred more adherence to the 3D scenes they created in terms of colour, lighting, and composition.

A central challenge lies in binding visual and textual inputs more effectively to provide better control. 
At the same time, concerns emerged around unpredictability. Creative video outputs were sometimes seen as ``too uncontrolled,'' straying from intended prompts or introducing misalignments between emotion, lighting, and movement. For instance, close-up reference videos conveyed facial emotion effectively, but these expressions could become diluted in wide-angle shots. Participants also noted that different genres demanded different motion strengths: while naturalistic drama favoured subtlety, animation often relied on exaggerated gestures to deliver narrative clarity. Addressing such mismatches requires systems that can adapt across modalities, detect inconsistencies (\eg, mood defined in text versus colour palette in visuals~\cite{yang2024emogen}), and adjust the material accordingly.  




\subsection{Assets Roughness vs. Iteration Cost}
In the current prototype, 3D assets can be incorporated at varying levels of fidelity, ranging from rough AI-generated objects to smoother, pre-defined geometries downloaded from external sources. These assets serve primarily as proxies within the environment. The degree of roughness has direct implications for refinement: smoother geometries often require fewer iterations to achieve a polished output, while rougher assets, particularly AI-generated characters, may demand more extensive refinement. For example, in our test sequence, a single refinement step was insufficient, whereas 19 iterations produced significantly improved results. This variation highlights the importance of \textit{asset quality} in shaping the efficiency of the refinement pipeline.

While these observations primarily highlight technical considerations for iterations using FlowEdit, they also offer design guidance. \textbf{Roughness} should not be seen solely as a limitation but as a feature that affords flexibility in early ideation. Designers and filmmakers can strategically decide when rough proxies are sufficient for blocking and experimentation, and when smoother assets are worth investing in to reduce refinement effort. 



\subsection{AI's Role in Shaping Collaborative Pre-production Through Previz}
Beyond individual authoring benefits, participants reflected on broader opportunities and limitations of AI-assisted previz. They saw value in using rough-to-polished workflows for communication and collaboration, especially when sharing ideas with clients or stakeholders who lacked technical or cinematic expertise. Early previews that combined structural clarity with stylistic polish helped avoid costly misunderstandings later in production.  


Taken together, these findings suggest that AI’s role in previz is not merely technical but collaborative~\cite{thomas2012collaboration}. Participants emphasized that roles and responsibilities shift depending on team size and production scale. In large teams, directors may rely on previz as a communication artifact across departments, while in smaller teams, AI-driven previz can extend into production itself, functioning as a lightweight substitute for traditional pipelines.  

By combining rough 3D environments with generative restyling and video remixing, \systemname allowed filmmakers to align creative intent across roles while accommodating diverse preferences for control versus creativity. 
This reflects AI’s potential to shape filmmaking practices not only in pre-production but across the broader ecosystem of creative collaboration~\cite{pal2025illuminating}.

\subsection{Ethical Consideration for Generative AI in Filmmaking}
Several participants raised concerns that point toward ethical considerations for future work~\cite{10.1145/3600211.3604681} (see \S\ref{sec:ai-creative-filmmaking}). While AI-driven previz enabled rapid iteration and creative flexibility, participants also noted risks around misrepresentation, unpredictability, and authorship. For example, outputs generated in more “creative” modes sometimes diverged significantly from intended prompts, producing movements or emotions that could mislead collaborators or clients during early decision-making. 

A related issue concerns the division of labour in filmmaking teams. Directors often focused on the polished outputs, delegating 3D scene setup to others, while costume designers, animators, and technical staff emphasized details such as colour palettes or fine-grained motion. By centralizing multiple creative functions into a single AI-driven workflow, systems like \systemname may blur traditional boundaries of responsibility. This raises questions about whose expertise is displaced, how credit should be attributed across roles, and whether generative previews risk oversimplifying or overwriting the specialized contributions of different departments 
~\cite{10.1145/3706598.3713342,hassapopoulou2024interactive}.  

Future work should therefore not only improve technical alignment across modalities but also embed safeguards for transparency, attribution, and consent in generative video workflows. 
For example, \systemname might expose the degree of resemblance versus creative divergence to make clear when outputs depart from the original scene, or provide provenance markers for external video inputs, and assist collaborative iteration to avoid misrepresenting creative intent.
\subsection{Limitation and Future Work}

Our study and prototype face several limitations. First, current generative models still struggle with controllability and continuity. \red{While character-specific LoRAs improved fidelity in close-up shots, they frequently degraded in wide-angle or occluded views, and scaling to multiple characters significantly slowed generation time.} Training such LoRAs also requires carefully curated datasets and caption labeling, which are uneven across styles and demographics, raising risks of bias. Continuity also breaks across props, lighting, and costumes, reducing coherence in multi-shot sequences, which often shifted across clips even under stricter adherence modes. \red{In addition, generative video models themselves occasionally introduce inconsistencies within a single clip, for example, adding unexpected visual elements or altering details between frames, which further challenges temporal stability. When non-character objects are important to the narrative or task (\eg, key props or in-scene advertising), dedicated LoRAs for these objects may also need to be trained to preserve their identity and appearance across shots; however, this level of fidelity is generally not essential for previsualization, as mentioned by participants, but may become important in real production contexts where object appearance must remain consistent.}
These factors limited coherence in multi-shot sequences and point to the need for more reliable mechanisms of cross-shot preservation. 
With the advances in AI and ML (\eg, GANs, and diffusion models), we expect these models and computation will become better in the future for creating realistic characters and consistent costumes~\cite{melki2019investigation,painguzhali2025artificial,zhang2024unlocking}, and more in terms of age, gender, and expression, and enable customizing facial features, hairstyles and clothing interactively~\cite{samy2025revolutionizing}. Future work could also address these challenges by reducing latency through progressive previews or asynchronous generation.  

Second, our observations stem from relatively short lab sessions (90–120 minutes across two scenes), which constrained the breadth of exploration and longer-form observations. Moreover, model latency limited rapid branching: in our setup, image/video generations typically required ~1 min per video clip, making it harder to explore wide parameter sweeps within a session. 
\red{We did not include a direct baseline comparison because existing previz or game-engine tools differ substantially in goals, interaction paradigms, and fidelity assumptions, making one-to-one comparisons difficult to interpret and easily confounded by factors such as modeling skill and asset variability. Instead, our focus was on understanding how practitioners work with a new unified workflow that combines 3D blocking, generative restyling, and motion conditioning. These findings can help inform future baseline-controlled studies by identifying which components, such as resemblance control, motion conditioning, or cross-shot continuity, are most consequential for practitioners and worth isolating in comparative and long-term evaluations.
Longer term field deployments with creative teams will be essential to understand how AI-drive previz intersects with production timelines, budgetary structures, and evolving divisions of labour in practice. }

\section{Conclusion}

We presented \systemname, a system that combines rough 3D scene blocking, detailed character motion, and video stylization through generative AI to support flexible, rapid previsualization. Through a user study with filmmakers, we found that the system enabled lightweight scene setup, iterative refinement, and expressive authoring across modalities. Our findings suggest that AI-assisted previz can augment creative practice, lowering barriers for independent creators to communicate their creative intent. At the same time, issues of latency, consistency, and fear of displacement highlight the need for careful future design.



\begin{acks}
We acknowledge the use of Cursor\footnote{https://cursor.com/} in the development of \systemname.
\end{acks}

\bibliographystyle{ACM-Reference-Format}
\bibliography{sample-base}

\appendix
\section{Appendix}
\label{appendix}
\red{To complement the scenes shown in the main paper, we provide
additional example outputs generated with \systemname{} to illustrate that
the workflow generalizes across a broader range of shot types (\autoref{sec:shotdiverse}) and film genres and visual styles (\autoref{sec:genre}) for previsualization. 
These examples were created using the same features described in
Sections~3--4 (3D blocking, resemblance-based restyling, and optional
video-driven motion) and do not represent additional case studies.
We include these to visualize the expressive range of the system.}

\subsection{\red{Shot Type Diversity}}
\label{sec:shotdiverse}
\begin{figure*}[ht]
    \centering
    \includegraphics[width=\linewidth]{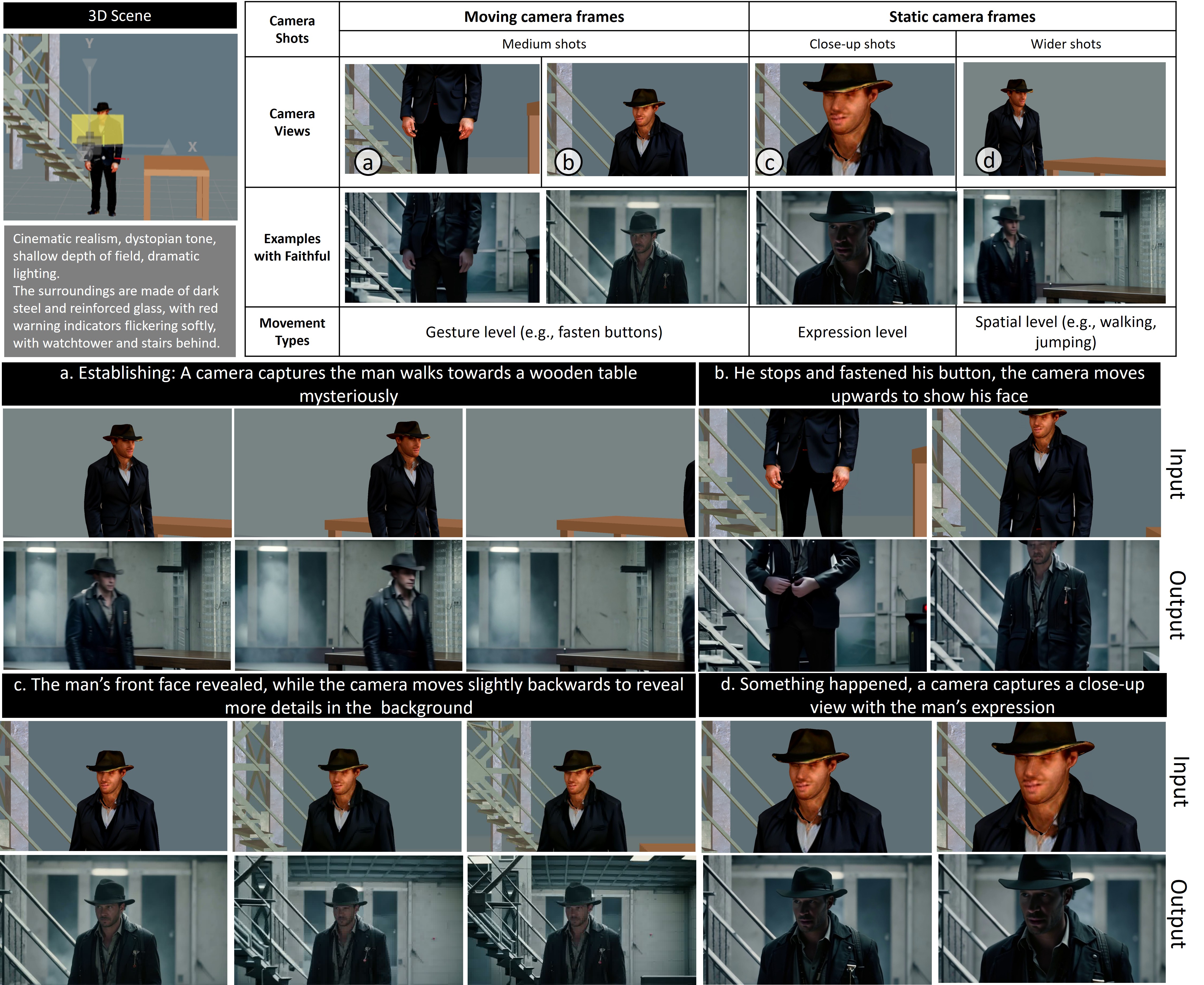}
    \caption{\red{Walkthrough of the “Mysterious Man Arrival” example. This figure illustrates how the system structures a simple narrative moment into two shots with distinct camera movement types.
Shot 1: (a–c) focuses on establishing spatial context and revealing character identity.
(a) Establishing movement: the camera observes the man as he walks toward a wooden table, introducing the environment and his mysterious presence.
(b) Upward reveal: as he stops and fastens his coat button, the camera tilts upward to gradually reveal his face.
(c) Backward drift: the camera subtly moves backward to expose more of the industrial background as the man’s full front view comes into frame.
Shot 2: (d) shifts to a close-up movement type.
(d) Emotional close-up: after an implied narrative beat, the camera cuts in to capture the man’s expression, emphasizing tension and character mood.}}
    \label{fig:expressions}
\end{figure*}

\begin{figure*}[ht]
    \centering
    \includegraphics[width=\linewidth]{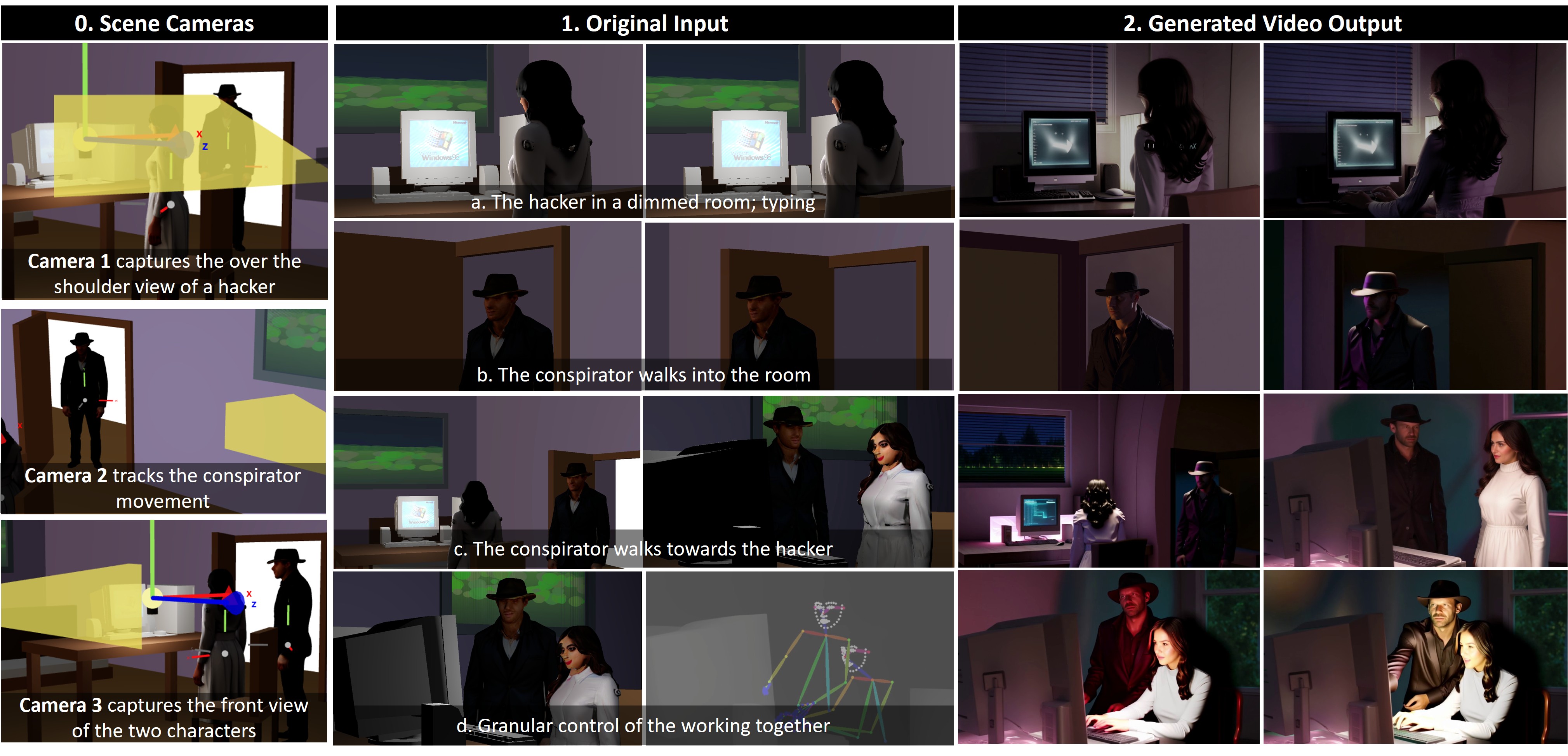}
    \caption{\red{Example 2: the “Hacker Scene” with original input (1) versus generated output (2).
(a) The hacker works alone in a dimly lit room, illuminated primarily by the glow of the computer screen. She types as the system’s 3D blocking panel defines the room layout, lighting, and character placement.
(b) The conspirator enters the room. The system updates lighting, occlusion, and character staging as he appears at the doorway and steps inside.
(c) The conspirator approaches the hacker. The system maintains spatial continuity and camera coherence as the distance between characters closes, showing how blocking, lighting, and viewpoint adjustments support tension-building character movement throughout the scene.}}
    \label{fig:app-shot}
\end{figure*}
\red{To demonstrate applicability to a wide range of camera practices,
\autoref{fig:expressions} and \autoref{fig:app-shot} shows outputs for different shot types commonly
used in filmmaking. Videos of these figure examples are provided in supplementary materials.} 






\red{For example, Example 1 (\autoref{fig:expressions}) illustrates how \systemname{} supports a range of standard filmmaking shots and camera movements within a single scene. 
The sequence begins with an establishing shot (a) showing the man walking toward the table, using wide framing to define space and character placement. It then transitions into a tracking push-in (b), as the camera moves upward to reveal the man’s face while he fastens his coat. Next, a medium shot (c) presents the character front-on, with a slight backward camera movement that exposes more background detail and maintains spatial continuity. Finally, a close-up reaction shot (d) captures the man’s expression, emphasizing emotional tone and character focus. Together, these stages show how \systemname handles different shot scales, character motions, and camera paths, supporting dynamic, narrative-driven coverage typical in cinematic previsualization.}

\subsection{Genre and Style Diversity}
\label{sec:genre}
\begin{figure*}[ht]
    \centering
    \includegraphics[width=\linewidth]{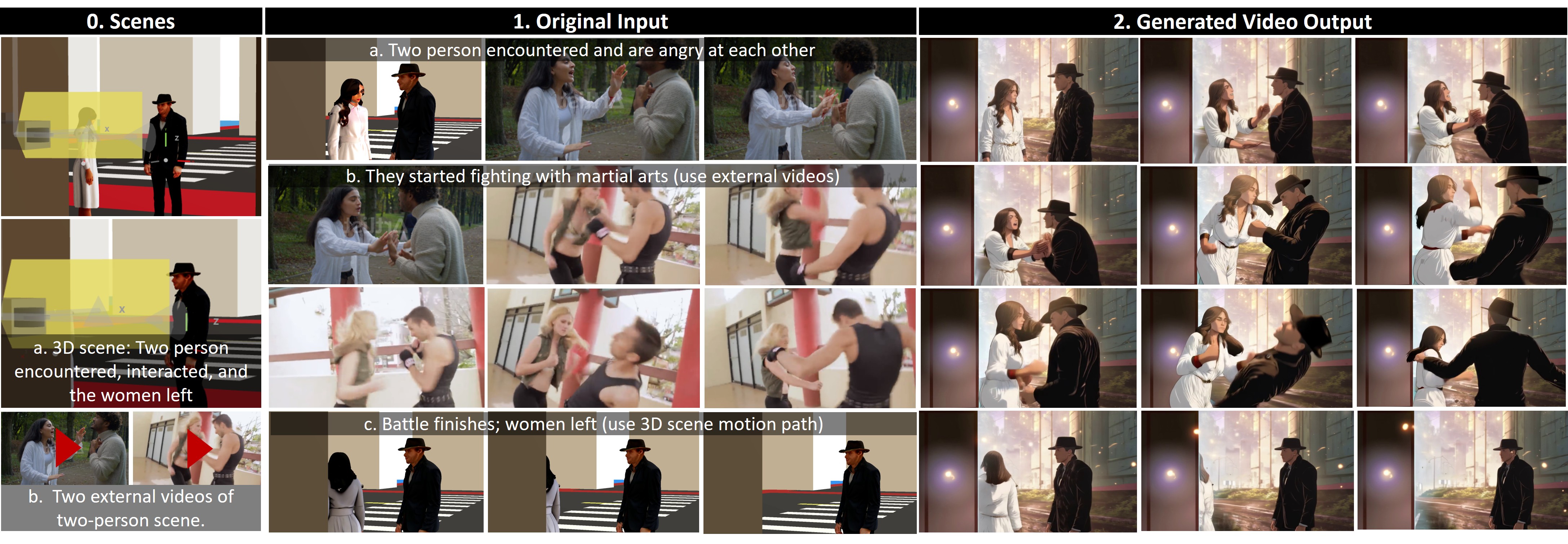}
    \caption{\red{Example 3: "Fighting Scene" (outdoor) generated with \systemname: (1) original input versus (2) generated output: (a) Two characters confront each other in a staged 3D scene, establishing tension and initial blocking. (b) A martial-arts fight sequence is generated by applying motion from external reference videos to the characters, enabling dynamic, high-energy interaction. (c) After the fight ends, the woman exits the scene along a 3D-defined motion path, demonstrating smooth integration between guided motion and scene layout.}}
    \label{fig:fighting}
\end{figure*}

\red{Our example scenes provided cinematic realistic, and various visual styles (\eg, 3D cartoon, anime, cinematic) that captures character movement, and interactions.  For example, the Hacker Scene example (\autoref{fig:app-shot}) reflects neo-noir and techno-thriller conventions with dimmed interiors, directional light from screens, high-contrast purples and blues, and tense two-character staging typical of surveillance, espionage, or cybercrime narratives.
In contrast, the Mysterious Man example (\autoref{fig:expressions}) adopts a detective or crime-drama visual language with muted industrial colours, trench-coat silhouettes, moody stairwell lighting, and cinematic close-ups emphasizing emotional ambiguity. Beyond subtle motions such as typing and walking transferred from external videos, \systemname can also reproduce high-energy action, such as martial-arts fighting sequences in Example 3 with Anime style (\autoref{fig:fighting}). These examples highlight that genre-specific cinematographic qualities can
be achieved via multimodal image-to-image and image-to-video models while still grounding compositions in the 3D scaffold.
}

\begin{table*}[]
\centering
\resizebox{\textwidth}{!}{%
\begin{tabular}{|c|c|c|c|c|c|c|c|}
\hline
\textbf{\begin{tabular}[c]{@{}c@{}}Participant\\ ID\end{tabular}} & \textbf{Age} & \textbf{Gender} & \textbf{Role} & \textbf{\begin{tabular}[c]{@{}c@{}}Year of \\ Experience\end{tabular}} & \textbf{AI during work} & \textbf{Description: Use of AI} & \textbf{Previz Experience} \\ \hline
1 & 24 & Female & \begin{tabular}[c]{@{}c@{}}Director/Writer,\\ Independent fictions\end{tabular} & 5 & Video Generation & \begin{tabular}[c]{@{}c@{}}Use AI generated videos as \\ characters’ dream/fantasy sequences\end{tabular} & \begin{tabular}[c]{@{}c@{}}Shot design and Storyboards\\  for breakdowns\end{tabular} \\ \hline
2 & 41 & Male & \begin{tabular}[c]{@{}c@{}}3D/Animation Experts \\ in filmmaking industry\end{tabular} & 12 & Image Generation & Image generation like MidJourney & Storyboards \\ \hline
3 & 30 & Female & \begin{tabular}[c]{@{}c@{}}Advertising Director\\ (Plot and documentary)\end{tabular} & 10 & N/A & N/A & \begin{tabular}[c]{@{}c@{}}Direct the staff to conduct tests \\ using storyboard or 3D previz\end{tabular} \\ \hline
4 & 30 & Male & Film students & 1 & \begin{tabular}[c]{@{}c@{}}Image / Video\\  Generation\end{tabular} & \begin{tabular}[c]{@{}c@{}}Expand the image for poster design; \\ use Video generation for scenery shots\end{tabular} & Cine Tracer \\ \hline
5 & 42 & Female & \begin{tabular}[c]{@{}c@{}}Styling and Costumes\\  in Filmmaking\end{tabular} & 15 & N/A & N/A & N/A \\ \hline
6 & 32 & Male & \begin{tabular}[c]{@{}c@{}}Indie technical director/\\ Cinematographer/\\  Content creator on short films\end{tabular} & 10 & \begin{tabular}[c]{@{}c@{}}Image / Video \\ Generation\end{tabular} & \begin{tabular}[c]{@{}c@{}}Use image models for reference; \\ Use video models (Runway, Veo, Kling) \\ to produce footages for indie /commercial videos\end{tabular} & \begin{tabular}[c]{@{}c@{}}Previs pro, Cine Tracer,\\  Shot Designer\end{tabular} \\ \hline
7 & 29 & Nonbinary & Art Handler / Art PA/Props & 6 & Image Generation & Creating systems to generate images with ComfyUI & \begin{tabular}[c]{@{}c@{}}Experience doing storyboards\\  and animatics. Both by hands \\ and using Harmony.\end{tabular} \\ \hline
8 & 29 & Male & \begin{tabular}[c]{@{}c@{}}Cinematographer for \\ Narrative films, documentaries, \\ social media content creators\end{tabular} & 7 & \begin{tabular}[c]{@{}c@{}}Image / Video\\ Generation\end{tabular} & \begin{tabular}[c]{@{}c@{}}Generating images on Midjourney, \\ especially with the intent of \\ creating pitch decks for films and series; \\ Video creation with Runway, Veo 2 (not extensively).\end{tabular} & Set a Light 3D, Cine Tracer \\ \hline
9 & 30 & Female & Brand advertising director, short films & 7 & \begin{tabular}[c]{@{}c@{}}Image/Video \\ Generation\end{tabular} & \begin{tabular}[c]{@{}c@{}}Involved in a project to\\  create short dramas using AI.\end{tabular} & \begin{tabular}[c]{@{}c@{}}Hire storyboard artists \\ to draw storyboards; \\ or use motion capture actors and \\ green screens to make previews.\end{tabular} \\ \hline
10 & 23 & Female & Technical Rigger and Animator & 2 & N/A & N/A & \begin{tabular}[c]{@{}c@{}}3d Previz/Layout, \\ 3d animation pipeline, \\ advertising, long features movies\end{tabular} \\ \hline
\end{tabular}%
}
\caption{Participant Demographics}
\label{tab:demographics}
\end{table*}

\end{document}